\documentclass[10pt,reprint,aps,prb]{revtex4-2}
\usepackage{amssymb}
\usepackage{graphicx}
\usepackage{multirow}
\usepackage{appendix}
\usepackage{mathtools}
\usepackage{amsmath}
\usepackage{xcolor}
\usepackage{siunitx}
\usepackage{yfonts}
\usepackage{esint}
\usepackage{placeins}
\graphicspath{{./IMG/}}

\newcommand{\TF}{\boldsymbol{\mathcal G}}

\usepackage{tikz}
\usetikzlibrary{arrows.meta,positioning}
\begin{document}

%\preprint{APS/123-QED}

\title{Global Buckley-Leverett theory for multicomponent flow in fractured media: Isothermal equation-of-state coupling and dynamic capillarity}

%\title{Global Buckley–Leverett Theory for Compositional Flow in Fractured Media}

\author{Christian Tantardini}
\email{christiantantardini@ymail.com}
\affiliation{Center for Integrative Petroleum Research, King Fahd University of Petroleum and Minerals, Dhahran 31261, Saudi Arabia}

\author{Fernando Alonso-Marroqu\'in}
\email{fernando@quantumfi.net}
\affiliation{Department of Computational Physics for Engineering Material, ETH Zurich, 8092 Zurich, Switzerland}
\affiliation{Center for Integrative Petroleum Research, King Fahd University of Petroleum and Minerals, Dhahran 31261, Saudi Arabia}

\date{\today}

\begin{abstract}
We present an isothermal Global Buckley--Leverett framework for multicomponent, multiphase flow in porous and fractured media that retains the interpretability of classical Buckley--Leverett while incorporating essential physics: equation of state-based phase behavior, multicomponent Maxwell--Stefan diffusion, dynamic capillarity, stress-sensitive permeability, and non-Darcy fracture flow. The formulation yields a single global-pressure equation driving the total Darcy flux and an exact fractional-flow decomposition of phase velocities with buoyancy and capillary drifts; inertial effects enter as per-phase damping that renormalizes mobilities. Crucially, the combination of Maxwell--Stefan diffusion and dynamic capillarity renders transport pseudo-parabolic, resolving the loss of strict hyperbolicity that plagues three-phase Buckley--Leverett and ensuring a well-posed initial-value problem. In practice, each time step solves the scalar global-pressure equation, reconstructs phase fluxes via the split, and advances strictly conservative component balances; axisymmetric (cylindrical) forms for radial injection with vertical buoyancy are provided. The model reduces exactly to classical Buckley--Leverett when added physics are disabled, making it a practical backbone for carbon storage, geothermal exchange, and contaminant transport in fractured, compositionally complex reservoirs.
\end{abstract}

\maketitle

\section{Introduction}
\label{sec:intro}

The Buckley--Leverett (BL) picture remains the simplest and most transparent way to reason about displacement in porous media. In its classical setting—two immiscible, incompressible phases obeying Darcy’s law with fixed pressure-volume-temperature (PVT) and negligible capillarity—the governing transport reduces to a scalar conservation law whose wave structure follows directly from the \emph{fractional-flow} curve, i.e., the mobility-weighted \emph{fraction} of the total flux carried by each phase. That blend of physical clarity and analytic control is why the original waterflood analysis by Buckley and Leverett, together with Leverett’s capillary scaling, still anchors pedagogy, verification, and quick-look analysis \citep{Buckley1942,Leverett1941}.

Difficulties arise as soon as we leave the two-phase world. For three immiscible phases, the transport system becomes two coupled conservation laws on the saturation simplex, and it is now well established that the model can \emph{lose strict hyperbolicity}. Eigenvalues coalesce along curves, umbilic points emerge, and parts of the state space effectively become elliptic; in those regions the textbook BL construction no longer selects a unique sequence of shocks and rarefactions unless additional physics is introduced to regularize the equations \citep{ShearerTrangenstein1989,Holden1990,Azevedo2010,Azevedo2014}. An elegant response is the global-pressure reformulation, which aims to decouple a single pressure equation from transport. For two phases the decoupling is fully equivalent; for three phases it is equivalent only when the data satisfy a stringent Total Differential (TD) compatibility among relative permeabilities and capillary pressures across the ternary diagram \citep{Chavent2009,DiChiara2010JCP}.
Throughout, ‘phases’ refers to fluid phases; the solid matrix is treated as an immobile skeleton that may store adsorbed mass.
These facts already hint that three-phase BL, taken literally, lacks stabilizing physics.

In unconventional (shale and ultra-tight) settings the missing mechanisms become decisive. The flow is often compositional and strongly compressible; phase changes; and components exchange across phases and with the solid via adsorption/desorption. Those effects lie outside the fixed-property, immiscible BL assumptions and instead call for an equation-of-state (EOS) framework that carries phase split, densities, and viscosities consistently \citep{Coats1980}. When composition varies, diffusion is inherently \emph{multicomponent} and cross-coupled: Maxwell--Stefan theory provides the thermodynamically consistent description and reduces to scalar Fickian diffusion only in special limits \citep{Krishna2014}. In nanoporous matrices, no-slip assumptions fail: gas slippage and molecule–wall collisions introduce Knudsen contributions, while adsorbed layers support surface diffusion along the solid \citep{Javadpour2009}. In fractures, connectivity and high rates make inertial (non-Darcy/Forchheimer) losses measurable, and both matrix permeability and fracture transmissivity evolve with effective stress \citep{Chen2015Forchheimer,Meng2020}. Finally, capillarity is rate dependent: there is strong theoretical and experimental evidence for \emph{dynamic} capillary pressure and the role of interfacial area, and—crucially for three-phase transport—those terms supply exactly the regularization that restores uniqueness where BL alone admits non-unique constructions \citep{HassanizadehGray1993,Bottero2011}.

Our aim is a mechanistic yet thermodynamically consistent BL-style formulation that integrates these ingredients into a single conservative transport system with an explicit fractional-flow split. We work isothermally (temperature is treated as a fixed parameter, as is standard in upstream applications) and present the equations in a coordinate-free way for generality. For analytical use in radial injection and buoyancy-dominated scenarios (e.g., CO$_2$ storage with axisymmetric drive and gravity along the axis), we also provide a concise specialization in cylindrical, axisymmetric coordinates later in the paper. 

This paper assembles a \emph{Global Buckley--Leverett for $N$ components and $N_p$ phases} (GBL-$N$) that: (i) keeps the BL interpretation visible through a phase-flux split; (ii) embeds EOS-consistent phase behavior, multicomponent diffusion, dynamic capillarity, non-Darcy fracture flow, stress sensitivity, and explicit fracture coupling; (iii) explains mathematically why classical three-phase BL can lose strict hyperbolicity and how Maxwell--Stefan diffusion and dynamic capillarity restore well-posedness; (iv) admits an exact global-pressure decoupling under TD (and a natural multi-phase extension, gTD), with a projection-based surrogate otherwise; and (v) reduces cleanly to classical BL when its assumptions hold. The operative system appears in \S\ref{subsec:gbln-operative}; the global-pressure structure is detailed in \S\ref{subsec:global-pressure}; the classical limit and well-posedness are summarized in \S\ref{subsec:classical-wellposed}; and fracture modeling is reviewed in \S\ref{subsec:fractures}.
The present formulation is a Darcy-scale model built on phase-based superficial velocities, relative permeabilities, and Darcy-type momentum closures (with a Forchheimer-type inertial correction used primarily in fractures). It is intended for regimes where Darcy-scale averaging remains meaningful (connected flow pathways and representative averaging) and where additional physics (EOS phase behavior, multicomponent diffusion, dynamic capillarity, stress sensitivity) is required.
We emphasize that multiphase flow may also exhibit nonlinear effective laws associated with pore-scale interface mobilization and intermittent pathway flow, often reported as a power-law relation between flow rate and pressure gradient; this regime is not captured by a per-phase Darcy plus Forchheimer closure \citep{Gao2020PRFluids,Zhang2021GRL}.
Accordingly, the present work does not claim to model that regime; rather, it provides a conservative thermodynamically consistent backbone into which alternative nonlinear Darcy-scale closures may be inserted when required (see also \S\ref{subsec:momentum}).

\section{Global Buckley--Leverett for multicomponent, multiphase flow}\label{sec:gbln}

This section assembles a multicomponent, multiphase Buckley--Leverett (GBL-$N$) formulation that preserves the BL intuition (a conservative transport system with a fractional-flow split) while adding only the physics needed in fractured, compositional settings. We work in conservative form throughout and make precise when a scalar global pressure exists.

\medskip

We consider \emph{isothermal} flow (temperature \(T\) is fixed). The spatial position is the vector \(\mathbf x\) (unrelated to the symbols \(x_{i\alpha}\), which denote \emph{phase mole fractions}). The overall composition is \(z=(z_1,\ldots,z_{N_c})\) with \(\sum_{i=1}^{N_c} z_i=1\).
Phases are indexed by \(\alpha=1,\dots,N_p\) (with only \(N_p^\star\!\le\!N_p\) present at equilibrium). We write mass density \(\rho_\alpha\) and molar density \(c_\alpha\) for each phase, linked by
\begin{align}
\bar M_\alpha = \sum_{i=1}^{N_c} M_i\,x_{i\alpha},
\qquad
\rho_\alpha = \bar M_\alpha\,c_\alpha ,
\label{eq:Mmix-relation}
\end{align}
where \(M_i\) is the molar mass of component \(i\).
We use \(\kappa\in\{\mathrm m,\mathrm f\}\) to label the \emph{matrix} (\(\mathrm m\)) and \emph{fracture} (\(\mathrm f\)) continua; if preferred, think of a boolean index with \(\kappa=0\) (matrix) and \(\kappa=1\) (fracture).

\subsection{State, closures, and thermodynamics}\label{subsec:state-thermo}

We adopt a minimal set of \emph{primitive variables} in each continuum \(\kappa\in\{\mathrm m,\mathrm f\}\):
pressure \(p\), (fixed) temperature \(T\), phase saturations \(S_\alpha^\kappa\) (with \(\sum_\alpha S_\alpha^\kappa=1\)), and a single \emph{overall} composition \(z=(z_1,\dots,z_{N_c})\) (with \(\sum_i z_i=1\)). Using overall composition rather than per-phase compositions is the classical “overall–composition” strategy: \(z\) is conserved across control volumes, evolves smoothly through \emph{phase transitions} (appearance/disappearance), and keeps transport strictly conservative \citep{Coats1980}.

Given \((p,T,z)\), an isothermal EOS flash returns the phase set and PVT data via a smooth map
\begin{align}
&(p,T,z)\ \longmapsto\ \big\{\nu_\alpha,\ x_{i\alpha},\ \rho_\alpha,\ \mu_\alpha\big\}_{\alpha=1}^{N_p^\star},
\label{eq:state-map}\\
&z_i = \sum_{\alpha=1}^{N_p^\star}\nu_\alpha\,x_{i\alpha},\qquad 
\sum_{\alpha=1}^{N_p^\star}\nu_\alpha = 1,\qquad 
\sum_{i=1}^{N_c}x_{i\alpha} = 1.
\label{eq:state-constraints}
\end{align}
Here \(\nu_\alpha\in[0,1]\) is the \emph{phase molar fraction} (so \(N_p^\star\) is the number of phases with \(\nu_\alpha>0\)), \(x_{i\alpha}\) is the mole fraction of component \(i\) in phase \(\alpha\), and \(\rho_\alpha,\mu_\alpha\) are phase density and viscosity.
While the EOS provides phase split and densities and enables the computation of chemical potentials (or fugacity coefficients) used for phase equilibrium, phase viscosities are obtained from separate transport-property correlations evaluated at the flashed state (e.g., Lohrenz--Bray--Clark and corresponding-states / EOS-based viscosity models)\citep{LohrenzBrayClark1964JPT,PedersenFredenslund1987CES,FanWang2006FPE,MichelsenMollerup2007,PrausnitzLichtenthalerAzevedo1999,MIT1040PRFugacity2003}.

At chemical equilibrium, for each component \(i\) there exists a scalar \(\Lambda_i\) such that the component fugacity is the same in all present phases:
\begin{align}
&\exists\,\{\Lambda_i\}_{i=1}^{N_c}\ \text{s.t.}\quad
f_i^\alpha(p,T,\boldsymbol x_\alpha) = \Lambda_i,
\label{eq:fugacity} \\
&\forall\,\alpha\ \text{with}\ \nu_\alpha>0,\ \ i=1,\dots,N_c. \nonumber
\end{align}
This is equivalent to “\(f_i^\alpha=f_i^\beta\)” for all present \(\alpha,\beta\), but avoids redundant \(O(N_p^2)\) pairwise constraints. Together with \eqref{eq:state-constraints}, it defines the flash. For cubic EOS Soave–Redlich–Kwong(SRK) / Peng–Robinson (PR),
\begin{align}
f_i^\alpha(p,T,\boldsymbol x_\alpha) \;=\; \varphi_{i\alpha}(p,T,\boldsymbol x_\alpha)\,x_{i\alpha}\,p,
\end{align}
where \(\varphi_{i\alpha}\) is the \emph{fugacity coefficient} (we use \(\varphi\) to avoid confusion with porosity \(\phi\)). For electrolyte/brine phases, an activity–coefficient model may replace \(\varphi_{i\alpha}\).
The relation \(f(p,T,\rho)=\text{const}\) applies to \emph{pure} fluids; for \emph{mixtures} the equilibrium condition is \eqref{eq:fugacity}—equality of \emph{each component’s} fugacity across phases \citep{Michelsen1982,MichelsenMollerup2007}. We do not commit to a specific EOS; we only require that the map \eqref{eq:state-map} be smooth and thermodynamically consistent \citep{Coats1980}.

Species drive one another; we model the diffusive \emph{molar} flux of component \(i\) in phase \(\alpha\) (relative to the phase-average velocity) as
\begin{align}
\mathbf J_{i\alpha} 
&= -\,c_\alpha \sum_{j=1}^{N_c}\mathbf D^{\mathrm{MS}}_{ij,\alpha}\cdot\nabla x_{j\alpha}
\;-\; \mathbf J_{i\alpha}^{\mathrm{Kn/surf}},
\label{eq:MSz}
\end{align}
with \(c_\alpha\) the phase molar density, \(\mathbf D^{\mathrm{MS}}_{ij,\alpha}\) a (possibly anisotropic) porous-media effective diffusivity, and \(\mathbf J_{i\alpha}^{\mathrm{Kn/surf}}\) a calibrated correction for slip/Knudsen/surface diffusion \citep{Javadpour2009}. By definition of the Maxwell--Stefan diffusive fluxes relative to the phase-average velocity, the diffusive molar fluxes satisfy
\begin{align}
\sum_{i=1}^{N_c}\mathbf J_{i\alpha} = \mathbf 0.
\label{eq:MSconstraint}
\end{align}
That is, $\mathbf J_{i\alpha}$ redistributes species \emph{within} phase $\alpha$ but does not carry net molar flux of the phase. Phase appearance/disappearance and adsorption/desorption modify storage and advective transport through the flashed state map $(p,T,z)\mapsto(\nu_\alpha,x_{i\alpha},c_\alpha,\rho_\alpha)$, but do not violate the within-phase Maxwell--Stefan constraint.

Equivalently, collecting $\mathbf J_{i\alpha}$ into $\mathbf J_\alpha=(\mathbf J_{1\alpha},\ldots,\mathbf J_{N_c\alpha})^\top$, one may write the Maxwell--Stefan closure in compact chemical--potential form as
\begin{align}
\mathbf J_\alpha &= -\,c_\alpha\,\mathbf D^{\mathrm{MS}}_\alpha\,\TF_\alpha\,\nabla \boldsymbol{x}_\alpha,
\label{eq:MSmu}
\\
[\TF_\alpha]_{ij} &= \frac{\partial \mu_{i\alpha}}{\partial \ln x_{j\alpha}},
\label{eq:MSmu-b}
\end{align}
where $\TF_\alpha$ is the thermodynamic-factor matrix and $\mathbf D^{\mathrm{MS}}_\alpha$ is the (effective) Maxwell--Stefan diffusivity operator; both are symmetric positive definite in stable phases \citep{Krishna2014}.
Here $\nabla\boldsymbol{x}_\alpha=(\nabla x_{1\alpha},\ldots,\nabla x_{N_c\alpha})^\top$, and the product $\mathbf D^{\mathrm{MS}}_\alpha\,\TF_\alpha\,\nabla\boldsymbol{x}_\alpha$ is understood as the standard contraction yielding the vector of spatial molar fluxes.
In anisotropic media, $\mathbf D^{\mathrm{MS}}_\alpha$ denotes a block operator acting on $\nabla\boldsymbol{x}_\alpha$; in isotropic form $\mathbf D^{\mathrm{MS}}_{ij,\alpha}=D^{\mathrm{MS}}_{ij,\alpha}\mathbf I$.

Intrinsic permeability \(k^\kappa\) and porosity \(\phi^\kappa\) may evolve with \emph{effective stress} \(\sigma'\). Exponential (log-linear) effective-stress relations are widely used as compact empirical fits for permeability (and, when needed, porosity) over practical stress ranges. Such forms are consistent with phenomenological crack-closure models (e.g., Gangi-type) and with laboratory trends in tight and fractured rocks \citep{ShiDurucan2016IJCG,ZhengZhengLiuJu2015IJRMMS,Walsh1981IJRMMS,Meng2020}. In this work, we therefore parameterize stress sensitivity via
\begin{align}
k^\kappa(p,\sigma') &= k_0^\kappa\,\exp\!\big[-\gamma_k^\kappa(\sigma'-\sigma_0')\big], \\
\phi^\kappa(p,\sigma') &= \phi_0^\kappa\,\exp\!\big[-\gamma_\phi^\kappa(\sigma'-\sigma_0')\big],
\label{eq:stress-laws}
\end{align}
so that permeability and porosity decrease monotonically with increasing effective stress. Here $(k_0^\kappa,\phi_0^\kappa)$ are reference values at $\sigma_0'$, and $\gamma_k^\kappa,\gamma_\phi^\kappa\ge 0$ are calibration coefficients obtained from laboratory measurements. Here $\sigma'$ denotes effective stress (e.g., $\sigma'=\sigma-bp$ with Biot coefficient $b\in[0,1]$). The effective stress follows Terzaghi/Biot, e.g.\ \(\sigma'=\sigma-b\,p\) with Biot coefficient \(b\in[0,1]\). More elaborate poroelastic couplings can replace \eqref{eq:stress-laws} without affecting the conservative transport structure \citep{Meng2020}.

Equilibrium capillary pressures \(p_{c,\alpha}(S)\) and relative permeabilities \(k_{r\alpha}(S)\) supply the static closures. When rates matter, we use the well-documented dynamic correction
\begin{align}
    p_{c,\alpha}^{\kappa,\mathrm{dyn}}(S_\alpha^\kappa)=p_{c,\alpha}^{\mathrm{AM}}(S_\alpha^\kappa)-\tau_\alpha^\kappa(S_\alpha^\kappa)\,\partial_t S_\alpha^\kappa,
\end{align}
where $\tau_\alpha^\kappa(S_\alpha^\kappa)$ ins the dynamic capillarity coefficient (units of $Pa \cdot s$), which later provides pseudo-parabolic regularization of transport.

We encode phase appearance/disappearance via nonnegative phase fractions and a stability indicator \(\psi_\alpha\) (e.g., a tangent-plane distance):
\begin{align}
\nu_\alpha \ge 0, \qquad \psi_\alpha(p,T,z) \ge 0, \qquad \nu_\alpha\,\psi_\alpha=0 .
\label{eq:complementarity}
\end{align}
This means: if a phase is \emph{present} (\(\nu_\alpha>0\)), it is exactly at stability limit (\(\psi_\alpha=0\)); if it is \emph{absent} (\(\nu_\alpha=0\)) the stability test may be strictly positive (\(\psi_\alpha>0\)). In practice we simply run the stability test and phase split (the “flash”) at each \((p,T,z)\); present phases satisfy \(\psi_\alpha=0\). We then transport \(z\) and recover \(\{\nu_\alpha,x_{i\alpha},\rho_\alpha,\mu_\alpha\}\) by flashing, which preserves conservation and smoothness across phase transitions \citep{Coats1980}.

\medskip
The state layer thus fixes the variables and closures we use below: the conserved unknowns are \(\{S_\alpha^\kappa\}\) and \(z\); the EOS flash \eqref{eq:state-map}–\eqref{eq:fugacity} supplies \(\{\nu_\alpha,x_{i\alpha},\rho_\alpha,\mu_\alpha\}\); multicomponent diffusion is in Maxwell--Stefan form \eqref{eq:MSz}–\eqref{eq:MSmu} with the constraint \eqref{eq:MSconstraint}; and stress enters \(k^\kappa,\phi^\kappa\) through \eqref{eq:stress-laws}. These ingredients are then combined with momentum, global pressure, and the fractional-flow split in the next subsections.

\subsection{Component conservation with adsorption and matrix--fracture exchange}\label{subsec:comp-balance}

Using the EOS map \eqref{eq:state-map}–\eqref{eq:fugacity} to recover \(\{c_\alpha^\kappa,\rho_\alpha^\kappa,\mu_\alpha^\kappa,x_{i\alpha}^\kappa\}\) from \((p,T,z)\), and the Maxwell–Stefan closures \eqref{eq:MSz}–\eqref{eq:MSmu} with the constraint \eqref{eq:MSconstraint}, the strictly conservative \emph{molar} balance of component \(i\) in continuum \(\kappa\in\{\mathrm m,\mathrm f\}\) is
\begin{align}
&\frac{\partial}{\partial t}
\Big[
\phi^\kappa \sum_{\alpha=1}^{N_p} S_\alpha^\kappa\, c_\alpha^\kappa\, x_{i\alpha}^\kappa
+\;
\rho_r^\kappa\, \Gamma_i^\kappa
\Big] \nonumber\\
&\quad+
\nabla\!\cdot
\Big[
\sum_{\alpha=1}^{N_p} c_\alpha^\kappa x_{i\alpha}^\kappa\,\mathbf v_\alpha^\kappa
\;-\;
\sum_{\alpha=1}^{N_p}\mathbf J_{i\alpha}^\kappa
\Big]
=
q_i^\kappa \;+\; T_i^{\mathrm{m}\leftrightarrow \mathrm{f}},
\label{eq:comp-balance-molar}
\end{align}
for \(i=1,\dots,N_c\). Here the first bracket is storage (mobile pore inventory plus adsorbed inventory on the solid, written with bulk rock density \(\rho_r^\kappa\)); the divergence acts on advective phase fluxes and on the multicomponent MS fluxes. Sources/sinks are \(q_i^\kappa\). The exchange \(T_i^{\mathrm m\leftrightarrow \mathrm f}\) couples matrix and fractures. \emph{Densities are not assumed constant}: both \(c_\alpha^\kappa\) and \(\rho_\alpha^\kappa\) are EOS functions of \((p,T,\boldsymbol x_\alpha)\).

If a mass-based statement is preferred, replace \(c_\alpha^\kappa x_{i\alpha}^\kappa\) by \(\rho_\alpha^\kappa w_{i\alpha}^\kappa\) with
\(
w_{i\alpha}^\kappa=\frac{M_i x_{i\alpha}^\kappa}{\bar M_\alpha^\kappa},\ 
\bar M_\alpha^\kappa=\sum_j M_j x_{j\alpha}^\kappa,\ 
\rho_\alpha^\kappa=\bar M_\alpha^\kappa c_\alpha^\kappa
\),
to obtain the mass-balance form exactly equivalent to \eqref{eq:comp-balance-molar}.

Adsorption relaxes toward an equilibrium isotherm driven by the same state \((p,T,z)\) used in the flash:
\begin{align}
\partial_t \Gamma_i^\kappa
=\frac{\Gamma_{i,\mathrm{eq}}^\kappa(p,T,z)-\Gamma_i^\kappa}{t_i^\kappa},
\label{eq:ads-kinetics}
\end{align}
with relaxation time \(t_i^\kappa\ge 0\). Since \(\Gamma_i^\kappa\) appears only in storage, it alters wave speeds without changing the conservative flux structure.

Intercontinuum transfer is phase specific and must include capillarity and gravity consistently. 
Define the phase potential (phase hydraulic potential)
\begin{align}
\Phi_\alpha^\kappa
:= p^\kappa - p_{c,\alpha}^{\kappa,\mathrm{dyn}} \;-\; \rho_\alpha^\kappa\,\mathbf g\!\cdot\!\mathbf x,
\label{eq:phase-potential}
\end{align}
and model matrix--fracture exchange in a dual-porosity (Warren--Root/Kazemi-type) form as a Darcy-like flux driven by the phase-potential difference \citep{WarrenRoot1963,Kazemi1976,GilmanKazemi1983}:
\begin{align}
\mathcal T_\alpha^{\mathrm{m}\leftrightarrow \mathrm{f}}
&=
\omega\, k_{\mathrm{int}}
\left(\frac{k_{r\alpha}}{\mu_\alpha}\right)^{\mathrm{up}}
\Big(\Phi_\alpha^{\mathrm f}-\Phi_\alpha^{\mathrm m}\Big),
\label{eq:WR-transfer}
\\
T_i^{\mathrm{m}\leftrightarrow \mathrm{f}}
&=
\sum_{\alpha=1}^{N_p}
\mathcal T_\alpha^{\mathrm{m}\leftrightarrow \mathrm{f}}\;
c_{\alpha}^{\mathrm{up}}\,x_{i\alpha}^{\mathrm{up}},
\label{eq:WR-component}
\end{align}
where $\omega$ is the (geometry-dependent) shape factor (units $1/\mathrm{m}^2$) and $k_{\mathrm{int}}$ is an interporosity (matrix-to-fracture) permeability scale. The mobility $\left(k_{r\alpha}/\mu_\alpha\right)^{\mathrm{up}}$ and the compositional factors $(c_\alpha x_{i\alpha})^{\mathrm{up}}$ are upwinded with respect to the sign of $\mathcal T_\alpha^{\mathrm{m}\leftrightarrow \mathrm{f}}$.
Using $\Phi_\alpha$ ensures the correct transfer direction when capillarity is strong and/or gravity opposes the imposed pressure gradient; in the limit of negligible capillarity and density contrasts, \eqref{eq:WR-transfer} reduces to a pressure-difference transfer law. We emphasize that we do not use the global pressure $p_g^\kappa$ in the transfer term because $p_g^\kappa$ is a mobility-weighted scalar that drives total flux within each continuum rather than a phase-specific potential.

For a matrix cell \(m\) and intersecting fracture segment \(f\), conservative non-neighbor connections (NNCs) use a phase-potential jump:
\begin{align}
F_{\alpha}^{m\to f}
&=
T_{mf}\,\lambda_{\alpha}^{mf}\,
\Big(\Phi_\alpha^{\mathrm f}-\Phi_\alpha^{\mathrm m}\Big),
\label{eq:NNC-phase}
\\
T_i^{m\to f} &= \sum_{\alpha=1}^{N_p} F_{\alpha}^{m\to f}\; c_{\alpha}^{\mathrm{up}}\;x_{i\alpha}^{\mathrm{up}}, \\
\lambda_{\alpha}^{mf}&=\frac{k_{r\alpha}^{\mathrm{up}}}{\mu_{\alpha}^{\mathrm{up}}},
\label{eq:NNC-component}
\end{align}
with \(T_{mf}\) the geometric transmissibility and “up” chosen by the sign of \(F_{\alpha}^{m\to f}\). Expanding \(\Phi_\alpha\) recovers the standard pressure, capillary, and gravity jumps used in projection-based embedded discrete fracture model (pEDFM), while strict conservation is preserved by equal-and-opposite interface fluxes.

\subsection{Momentum with non-Darcy and dynamic capillarity}\label{subsec:momentum}

We close the phase velocities with a Darcy backbone plus three effects that are essential in modern fractured settings: (i) inertial corrections (Forchheimer) in fractures and other high-rate/rough pathways, (ii) stress-sensitive permeability/porosity \eqref{eq:stress-laws}, and (iii) a pore-morphology-derived equilibrium capillary law augmented by a dynamic (rate-dependent) term. For each phase \(\alpha\) in continuum \(\kappa\in\{\mathrm m,\mathrm f\}\),
\begin{align}
&-\nabla\!\Big(p - \rho_\alpha^\kappa\,\mathbf g\!\cdot\!\mathbf x\Big) \nonumber \\
&-\nabla p_{c,\alpha}^{\kappa,\mathrm{dyn}}(S_\alpha^\kappa)
=
\frac{\mu_\alpha^\kappa}{k^\kappa(p,\sigma')\,k_{r\alpha}^\kappa(S)}\,\mathbf v_\alpha^\kappa 
+
\beta_\alpha^\kappa\,\rho_\alpha^\kappa\,\|\mathbf v_\alpha^\kappa\|\,\mathbf v_\alpha^\kappa,
\label{eq:momentum-nd}
\end{align}
where $k^\kappa(p,\sigma')$ is the stress-sensitive intrinsic permeability and $k_{r\alpha}^\kappa(S)$ is the relative permeability. The Forchheimer coefficient $\beta_\alpha^\kappa$ is treated as a calibrated closure: in fractures it may depend on geometry (hydraulic aperture and roughness) and thus can be stress dependent through $b(\sigma')$, whereas in the matrix we typically set $\beta_\alpha^{\mathrm m}=0$ unless core data indicate otherwise.
The Forchheimer term here is used as a pragmatic inertial damping for high-rate fracture flow; it does not represent interface-mobilization-induced nonlinear multiphase rheology (often manifesting as power-law Darcy-scale behavior) \citep{Gao2020PRFluids}.
When needed, the framework admits alternative nonlinear Darcy-scale closures by replacing the per-phase momentum relation with a generalized mobility operator, e.g.,
\(
\mathbf v_\alpha^\kappa =
-\,k^\kappa\,\lambda_\alpha^\kappa\,
\mathcal N_\alpha\!\big(|\nabla \Phi_\alpha^\kappa|\big)\,
\nabla \Phi_\alpha^\kappa,
\)
without changing the conservative component balances or the global-pressure/fractional-flow structure \citep{Gao2020PRFluids,Zhang2021GRL}.

The \emph{dynamic} capillary pressure is decomposed into an equilibrium contribution and a rate-dependent correction,
\begin{align}
p_{c,\alpha}^{\kappa,\mathrm{dyn}}(S_\alpha^\kappa)
&=
p_{c,\alpha}^{\mathrm{AM}}(S_\alpha^\kappa;\gamma_{\alpha},\theta_{\alpha},\mathcal M_\kappa,\xi_\kappa)
-
\tau_\alpha^\kappa(S_\alpha^\kappa)\,\partial_t S_\alpha^\kappa .
\label{eq:pc-dynamic}
\end{align}

The Alonso--Marroquín \& Andersson pore-morphology \citep{AlonsoMarroquinAndersson2025PMM} method maps saturation to an effective \emph{capillary radius} selected by invasion under Young–Laplace balance. A quasi-2D curvature gives
\begin{align}
r_c \;=\; \frac{\gamma_{\alpha}\,\cos\theta_{\alpha}}{\,p_c^{\mathrm{eff}}\,},
\label{eq:AM-rc}
\end{align}
so a prescribed \(p_c^{\mathrm{eff}}\) selects all throats with \(r\le r_c\) as potentially invaded. Advancing invasion by connectivity operations (with a trapped set) returns \(S\) for each \(r_c\); inversion yields the equilibrium law
\begin{align}
p_{c,\alpha}^{\mathrm{AM}}(S;\gamma_{\alpha},\theta_{\alpha},\mathcal M_\kappa,\xi_\kappa)
\;=\;
\kappa_d\,
\frac{\gamma_{\alpha}\,\cos\theta_{\alpha}}{\,r_c\!\left(S;\,\mathcal M_\kappa,\xi_\kappa\right)},
\label{eq:AM-pcS}
\end{align}
with \(\kappa_d=1\) for the effective 2D curvature (and \(\kappa_d=2\) for axisymmetric cylinders).

Two routes are practical: (i) \emph{image-informed}—infer the throat-radius cumulative distribution function (CDF) \(F_r(r)\) from micro-CT or mercury intrusion, define an effective saturation \(S_e\) (accounting for residuals and trapping \(\xi_\kappa\)), set \(r_c(S)=F_r^{-1}(1-S_e)\), and tabulate \(p_{c,\alpha}^{\mathrm{AM}}(S)\) from \eqref{eq:AM-pcS}; (ii) \emph{surrogate fit}—when detailed morphology is unavailable, fit a monotone surrogate (e.g., Brooks--Corey, van Genuchten, or a smooth spline) to measurements and interpret it as \(p_{c,\alpha}^{\mathrm{AM}}(S)\). The continuum model needs only a smooth \(p_c(S)\); the morphology furnishes a constructive path and hysteresis via \(\xi_\kappa\).

Substituting \eqref{eq:pc-dynamic} into \eqref{eq:momentum-nd} gives
\begin{align}
&-\nabla\Big(p - \rho_\alpha^\kappa \mathbf g\!\cdot\!\mathbf x\Big)
\;-\;
\nabla p_{c,\alpha}^{\mathrm{AM}}(S_\alpha^\kappa;\cdots) \nonumber \\
&+
\tau_\alpha^\kappa(S_\alpha^\kappa)\,\nabla\!\big(\partial_t S_\alpha^\kappa\big) 
=
\frac{\mu_\alpha^\kappa}{k^\kappa k_{r\alpha}^\kappa}\,\mathbf v_\alpha^\kappa
\;+\;
\beta_\alpha^\kappa\,\rho_\alpha^\kappa\,\|\mathbf v_\alpha^\kappa\|\,\mathbf v_\alpha^\kappa .
\label{eq:momentum-expanded}
\end{align}
The gradient of \(p_{c,\alpha}^{\mathrm{AM}}(S)\) couples momentum to \(\nabla S\) via \(\partial p_c^{\mathrm{AM}}/\partial S\); the dynamic term produces the \emph{pseudo-parabolic} regularization used in the energy estimate (see \S\ref{subsec:classical-wellposed}).

In fractures, inertial departures from Darcy are \emph{not optional}: rough-walled data show measurable non-Darcy losses at modest in-plane Reynolds numbers, with losses increasing as aperture decreases \citep{Chen2015JHydrol}. We therefore take \(\beta_\alpha^{\mathrm f}>0\) by default (calibrated from fracture tests) and \(\beta_\alpha^{\mathrm m}=0\) in matrix unless core data indicate otherwise. A convenient dimensionless trigger for significance is
\begin{align}
\mathcal I_\alpha^\kappa
:= \beta_\alpha^\kappa\,\rho_\alpha^\kappa\,u_{\alpha}^{\kappa,D}\,
\frac{k^\kappa k_{r\alpha}^\kappa}{\mu_\alpha^\kappa},
\end{align}
where \(u_{\alpha}^{\kappa,D}=\|\mathbf v_{\alpha}^{\kappa,D,\Pi}\|\) is the Darcy driving speed defined in \eqref{eq:darcy-driving}. For \(\mathcal I_\alpha^\kappa\gtrsim 0.1\), the exact damping \(\chi_\alpha^\kappa\) from the positive root of the Forchheimer quadratic (eq.~\eqref{eq:chi-factor}) departs appreciably from unity; for \(\mathcal I_\alpha^\kappa\ll 0.1\), the linearization \eqref{eq:chi-linear} suffices. On fracture planes, an equivalent criterion uses the in-plane Reynolds number \(\mathrm{Re}_\tau^\alpha=(\rho_\alpha^{\mathrm f}\,u_{\alpha}^{\mathrm f,D}\,b)/\mu_\alpha^{\mathrm f}\), with onset typically \(\mathrm{Re}_\tau^\alpha\sim 1\!-\!10\) depending on roughness; empirically \(\beta_\alpha^{\mathrm f}\) scales like \(C_r/b\) with a roughness constant \(C_r\). In all cases we compute \(\chi_\alpha^\kappa\) via \eqref{eq:chi-factor} and use it consistently in the fractional-flow split and at matrix–fracture interfaces (cf.\ \S\ref{subsec:ff-split}, \S\ref{subsec:fractures}).

Densities \(\rho_\alpha\) appear \emph{explicitly} in momentum \eqref{eq:momentum-nd}–\eqref{eq:momentum-expanded} and \emph{implicitly} in transport through the EOS relation \(\rho_\alpha=\bar M_\alpha c_\alpha\) (cf.\ \eqref{eq:Mmix-relation}); no constant-density assumption is required. The capillary law \(p_{c,\alpha}^{\mathrm{AM}}(S)\) can be image-informed or a smooth surrogate; dynamic capillarity supplies the pseudo-parabolic term needed for well-posedness, while Forchheimer losses are retained where they matter most (fractures) and reduce to Darcy when inactive.

\subsection{Buckley--Leverett–style fractional–flow split}\label{subsec:ff-split}

Our objective is to retain the Buckley--Leverett intuition—each \emph{phase} flux is a mobility–weighted share of the total plus well–defined drifts—while remaining exact with the momentum law \eqref{eq:momentum-nd} and the dynamic capillarity \eqref{eq:pc-dynamic}. All expressions are coordinate–free (valid in Cartesian or cylindrical frames); gravity enters through the body–force form \(\rho_\alpha^\kappa\mathbf g\).

Define phase mobilities, total mobility, and the (actual) total superficial velocity
\begin{align}
\lambda_\alpha^\kappa(S) &= \frac{k_{r\alpha}^\kappa(S)}{\mu_\alpha^\kappa}, 
&
\Lambda_t^\kappa &= \sum_{\beta=1}^{N_p}\lambda_\beta^\kappa, 
&
\mathbf v_t^\kappa &= \sum_{\beta=1}^{N_p}\mathbf v_\beta^\kappa .
\label{eq:mob-totals}
\end{align}
Here, $\mathbf v_\alpha$ denotes the phase superficial (Darcy) velocity (volumetric flux per bulk cross-sectional area) \citep{Bear1972,Stauffer2006Flux}.
The total superficial velocity is defined as $\mathbf v_t=\sum_\alpha \mathbf v_\alpha$ \citep{Peaceman1977,AzizSettari1979}.
Pore (seepage) velocities are obtained by dividing by $\phi S_\alpha$ if needed \citep{Bear1972}.
To avoid a clash with the \emph{fugacity} notation \(f_i^\alpha\) used earlier, we denote the fractional–flow weights by
\[
F_\alpha^\kappa := \frac{\lambda_\alpha^\kappa}{\Lambda_t^\kappa}.
\]

Set \(\beta_\alpha^\kappa=0\) in \eqref{eq:momentum-nd}. Using \eqref{eq:pc-dynamic},
\begin{align}
\mathbf v_\alpha^{\kappa,D}
&= -\,k^\kappa\,\lambda_\alpha^\kappa\,
\Big(\nabla p \;-\; \rho_\alpha^\kappa\mathbf g \;-\; \nabla p_{c,\alpha}^{\kappa,\mathrm{dyn}}\Big), \\
p_{c,\alpha}^{\kappa,\mathrm{dyn}}
&=
p_{c,\alpha}^{\mathrm{AM}}(S_\alpha^\kappa;\cdots)
\;-\;
\tau_\alpha^\kappa(S_\alpha^\kappa)\,\partial_t S_\alpha^\kappa .
\label{eq:phase-Darcy}
\end{align}
Summing over phases gives
\begin{align}
\mathbf v_t^{\kappa,D}
=& -\,k^\kappa
\Big(
\Lambda_t^\kappa\,\nabla p 
\;-\; \sum_{\beta}\lambda_\beta^\kappa\,\rho_\beta^\kappa\,\mathbf g \nonumber \\
&- \sum_{\beta}\lambda_\beta^\kappa\,\nabla p_{c,\beta}^{\kappa,\mathrm{dyn}}
\Big).
\label{eq:total-Darcy}
\end{align}
Eliminating \(\nabla p\) between \eqref{eq:phase-Darcy} and \eqref{eq:total-Darcy} yields the \emph{exact} Darcy-level identity
\begin{align}
\mathbf v_\alpha^{\kappa,D}
= 
\underbrace{\frac{\lambda_\alpha^\kappa}{\Lambda_t^\kappa}}_{F_\alpha^\kappa}
\mathbf v_t^{\kappa,D}
&+
k^\kappa\,\lambda_\alpha^\kappa\,
\Big[
\big(\rho_\alpha^\kappa-\bar\rho_\lambda^\kappa\big)\,\mathbf g \nonumber \\
&+
\big(\nabla p_{c,\alpha}^{\kappa,\mathrm{dyn}}-\overline{\nabla p_c}^{\,\kappa}\big)
\Big],
\label{eq:darcy-split}
\end{align}
with mobility–weighted means
\begin{align}
\bar\rho_\lambda^\kappa
&=\frac{1}{\Lambda_t^\kappa}\sum_{\beta}\lambda_\beta^\kappa\,\rho_\beta^\kappa, \\
\overline{\nabla p_c}^{\,\kappa}
&=\frac{1}{\Lambda_t^\kappa}\sum_{\beta}\lambda_\beta^\kappa\,\nabla p_{c,\beta}^{\kappa,\mathrm{dyn}}.
\label{eq:lambda-averages}
\end{align}
The first term is the BL core \(F_\alpha^\kappa\,\mathbf v_t^{\kappa,D}\). The bracket collects \emph{drifts}: a buoyancy drift from \(\rho_\alpha^\kappa-\bar\rho_\lambda^\kappa\) and a capillary drift from departures relative to \(\overline{\nabla p_c}^{\,\kappa}\). Because \(p_{c,\alpha}^{\kappa,\mathrm{dyn}}=p_{c,\alpha}^{\mathrm{AM}}(S)-\tau_\alpha^\kappa(S)\partial_t S_\alpha^\kappa\), the capillary drift contributes both a standard second–order smoothing \((\partial_S p_c^{\mathrm{AM}}\,\nabla S)\) and a pseudo–parabolic term \((\nabla\partial_t S)\).

Return to \eqref{eq:momentum-nd} with \(\beta_\alpha^\kappa\!\neq\!0\). Define the Darcy driving vector and its magnitude,
\begin{align}
\mathbf v_{\alpha}^{\kappa,D,\,\Pi}
&\equiv
-\,\frac{k^\kappa k_{r\alpha}^\kappa}{\mu_\alpha^\kappa}\,
\Big(\nabla p \;-\; \rho_\alpha^\kappa\mathbf g \;-\; \nabla p_{c,\alpha}^{\kappa,\mathrm{dyn}}\Big),
\label{eq:darcy-driving}
\end{align}
with \(u_{\alpha}^{\kappa,D}=\|\mathbf v_{\alpha}^{\kappa,D,\,\Pi}\|\).
The Forchheimer relation gives a scalar quadratic for the actual speed \(u_\alpha^\kappa\):
\begin{align}
u_{\alpha}^{\kappa,D}
&= u_\alpha^\kappa \;+\; \beta_\alpha^\kappa\,\rho_\alpha^\kappa\,
\frac{k^\kappa k_{r\alpha}^\kappa}{\mu_\alpha^\kappa}\,\big(u_\alpha^\kappa\big)^2,
\label{eq:forchheimer-quadratic}
\end{align}
whose positive root defines a damping factor \(\chi_\alpha^\kappa\in(0,1]\) such that
\begin{align}
\mathbf v_\alpha^\kappa
&= \chi_\alpha^\kappa\,\mathbf v_{\alpha}^{\kappa,D,\,\Pi}, \\
\chi_\alpha^\kappa
&=
\frac{-1 + \sqrt{\,1 + 4\,\beta_\alpha^\kappa\,\rho_\alpha^\kappa\,
\frac{k^\kappa k_{r\alpha}^\kappa}{\mu_\alpha^\kappa}\,u_{\alpha}^{\kappa,D}\,}}{2\,\beta_\alpha^\kappa\,\rho_\alpha^\kappa\,
\frac{k^\kappa k_{r\alpha}^\kappa}{\mu_\alpha^\kappa}\,u_{\alpha}^{\kappa,D}}.
\label{eq:chi-factor}
\end{align}
In the weakly inertial regime,
\begin{align}
\chi_\alpha^\kappa \;\approx\; 1 \;-\;
\beta_\alpha^\kappa\,\rho_\alpha^\kappa\,
\frac{k^\kappa k_{r\alpha}^\kappa}{\mu_\alpha^\kappa}\,u_{\alpha}^{\kappa,D},
\label{eq:chi-linear}
\end{align}
consistent with rough–walled fracture data. Applying \(\chi_\alpha^\kappa\) to \eqref{eq:darcy-split} gives
\begin{align}
\mathbf v_\alpha^\kappa
&=
\Bigg\{
F_\alpha^\kappa\,\mathbf v_t^{\kappa,D}
\;+\;
k^\kappa\,\lambda_\alpha^\kappa\!
\Big[
\big(\rho_\alpha^\kappa-\bar\rho_\lambda^\kappa\big)\,\mathbf g \nonumber \\
&\qquad+
\big(\nabla p_{c,\alpha}^{\kappa,\mathrm{dyn}}-\overline{\nabla p_c}^{\,\kappa}\big)
\Big]
\Bigg\}\,\chi_\alpha^\kappa .
\label{eq:full-split}
\end{align}

With damping, the true total is \(\mathbf v_t^\kappa=\sum_\alpha \mathbf v_\alpha^\kappa \neq \mathbf v_t^{\kappa,D}\) (each phase has its own \(\chi_\alpha^\kappa\)). To keep a BL–style transport form, introduce \emph{apparent} mobilities and weights
\begin{align}
\tilde{\lambda}_\alpha^\kappa &\equiv \chi_\alpha^\kappa\,\lambda_\alpha^\kappa, \\
\tilde{\Lambda}_t^\kappa &\equiv \sum_{\beta}\tilde{\lambda}_\beta^\kappa, \\
\tilde{F}_\alpha^\kappa &\equiv \frac{\tilde{\lambda}_\alpha^\kappa}{\tilde{\Lambda}_t^\kappa},
\label{eq:apparent-mobs}
\end{align}
and write the \emph{operational} split
\begin{align}
\mathbf v_\alpha^\kappa
&=
\tilde F_\alpha^\kappa\,\mathbf v_t^\kappa
\;+\;
k^\kappa\,\tilde{\lambda}_\alpha^\kappa
\Big[
\big(\rho_\alpha^\kappa-\tilde{\bar\rho}_\lambda^\kappa\big)\,\mathbf g \nonumber \\
&\qquad+
\big(\nabla p_{c,\alpha}^{\kappa,\mathrm{dyn}}-\widetilde{\overline{\nabla p_c}}^{\,\kappa}\big)
\Big],
\label{eq:operational-split}\\
\tilde{\bar\rho}_\lambda^\kappa
&=\frac{1}{\tilde{\Lambda}_t^\kappa}\sum_{\beta}\tilde{\lambda}_\beta^\kappa\,\rho_\beta^\kappa, \\
\widetilde{\overline{\nabla p_c}}^{\,\kappa}
&=\frac{1}{\tilde{\Lambda}_t^\kappa}\sum_{\beta}\tilde{\lambda}_\beta^\kappa\,\nabla p_{c,\beta}^{\kappa,\mathrm{dyn}}.
\label{eq:apparent-averages}
\end{align}
An overbar $\bar{(\cdot)}_\lambda$ denotes a mobility-weighted average using $\lambda_\alpha$; a tilde $\tilde{(\cdot)}$ denotes Forchheimer-damped (‘apparent’) quantities using $\tilde\lambda_\alpha=\chi_\alpha\lambda_\alpha$; superscript D denotes the Darcy-level (no inertia) flux.
Thus Forchheimer effects enter as colinearity damping folded into \(\tilde{\lambda}_\alpha^\kappa\). No global pressure is assumed in this derivation; when the TD/gTD condition holds, the same split coexists with a cleaner pressure equation (see \S\ref{subsec:global-pressure}).

\subsection{Global pressure and the generalized TD hypothesis}\label{subsec:global-pressure}

The role of a global pressure is to collect mobility–weighted \emph{phase–pressure} gradients into a \emph{single} scalar gradient so that, at the Darcy level, the total flux is driven by one unknown.  
Consistently with \eqref{eq:momentum-nd} and the convention \(p_\alpha:=p-p_{c,\alpha}^{\kappa,\mathrm{dyn}}\) (with \(p_{c,\alpha}^{\kappa,\mathrm{dyn}}\) from \eqref{eq:pc-dynamic}), we \emph{define} \(p_g^\kappa\) pointwise by
\begin{align}
\sum_{\alpha=1}^{N_p}\lambda_\alpha^\kappa(S)\,\big(\nabla p_\alpha^\kappa - \rho_\alpha^\kappa \mathbf g\big)
= \Lambda_t^\kappa(S)\,\nabla p_g^\kappa .
\label{eq:pg}
\end{align}
This identity is coordinate–free (Cartesian or cylindrical). When \(\rho_\alpha^\kappa\) varies spatially, the body–force form \(-\nabla p_\alpha^\kappa+\rho_\alpha^\kappa\mathbf g\) avoids spurious \(\nabla\rho_\alpha^\kappa\) terms.

Because \(p_\alpha=p-p_{c,\alpha}^{\kappa,\mathrm{dyn}}\), \eqref{eq:pg} implies
\begin{align}
\Lambda_t^\kappa\,\nabla p
= \Lambda_t^\kappa\,\nabla p_g^\kappa
+ \sum_{\beta=1}^{N_p}\lambda_\beta^\kappa\,\nabla p_{c,\beta}^{\kappa,\mathrm{dyn}}
+ \sum_{\beta=1}^{N_p}\lambda_\beta^\kappa\,\rho_\beta^\kappa\,\mathbf g,
\label{eq:nabla-p-from-pg}
\end{align}
i.e.
\(
\nabla p
= \nabla p_g^\kappa + \overline{\nabla p_c}^{\,\kappa} + \bar\rho_\lambda^\kappa\,\mathbf g
\)
with the mobility–weighted means in \eqref{eq:lambda-averages}.  
Substituting \eqref{eq:nabla-p-from-pg} into the Darcy expressions yields the \emph{Darcy total flux}
\begin{align}
\mathbf v_t^{\kappa,D} 
= -\,k^\kappa\,\Lambda_t^\kappa\,\nabla p_g^\kappa,
\label{eq:vt-gp}
\end{align}
and the \emph{Darcy fractional–flow split} \eqref{eq:darcy-split}, \emph{without computing \(p\) or \(p_\alpha\) explicitly}. In practice, we solve for \(p_g^\kappa\) and then compute \(\mathbf v_t^{\kappa,D}\) and the phase fluxes via \eqref{eq:darcy-split} (or the operational split \eqref{eq:operational-split} when Forchheimer damping is active).

In cylindrical coordinates with no $\theta$–dependence, replace $\nabla$ and $\nabla\!\cdot$ by their axisymmetric forms as in §\ref{subsec:axisym}; in particular,
\(
-\nabla\!\cdot(k^\kappa\Lambda_t^\kappa\nabla p_g^\kappa)
= -\frac{1}{r}\partial_r\!\big(r\,k^\kappa\Lambda_t^\kappa\,\partial_r p_g^\kappa\big)
-\partial_z\!\big(k^\kappa\Lambda_t^\kappa\,\partial_z p_g^\kappa\big).
\)

For \(N_p=2\) with equilibrium \(p_c(S)\), \eqref{eq:pg}–\eqref{eq:vt-gp} reproduce the classical global–pressure/fractional–flow split: capillarity enters transport as drifts (cf.\ \eqref{eq:darcy-split}), while \(\mathbf v_t^{\kappa,D}\) is driven solely by \(\nabla p_g^\kappa\).

For \(N_p=3\), ask whether the mobility–weighted \emph{equilibrium} capillary term defines a saturation potential:
\begin{align}
\frac{1}{\Lambda_t^\kappa(S)}\sum_{\alpha=1}^3\lambda_\alpha^\kappa(S)\,\nabla p_{c,\alpha}^\kappa(S)
\overset{?}{=}\nabla \Pi^\kappa(S_1,S_2).
\label{eq:TD-target}
\end{align}
This holds \emph{iff} the 1–form
\begin{align}
    \eta^\kappa(S)=\frac{1}{\Lambda_t^\kappa(S)}\sum_{\alpha=1}^3 \lambda_\alpha^\kappa(S)\, d p_{c,\alpha}^\kappa(S)
\end{align}
is exact on the ternary simplex, i.e.\ the TD compatibility \citep{Chavent2009,DiChiara2010JCP}:
\begin{align}
\frac{\partial}{\partial S_2}
\Bigg[
\frac{1}{\Lambda_t^\kappa}\sum_{\alpha=1}^3\lambda_\alpha^\kappa\,
\frac{\partial p_{c,\alpha}^\kappa}{\partial S_1}
\Bigg]
=
\frac{\partial}{\partial S_1}
\Bigg[
\frac{1}{\Lambda_t^\kappa}\sum_{\alpha=1}^3\lambda_\alpha^\kappa\,
\frac{\partial p_{c,\alpha}^\kappa}{\partial S_2}
\Bigg].
\label{eq:TD-condition}
\end{align}
When \eqref{eq:TD-condition} holds, a potential \(\Pi^\kappa\) exists (unique up to a constant), and the global–pressure/transport decoupling is fully equivalent to the original three–phase equations; then \(\nabla p_g^\kappa=\nabla p-\nabla \Pi^\kappa-\bar\rho_\lambda^\kappa\mathbf g\).

Let \(s=(S_1,\dots,S_{N_p-1})\) be independent saturations on the simplex. Define the \emph{mobility–weighted capillary field}
\begin{align}
\mathbf A^\kappa(s)
&:=\big(A_1^\kappa(s),\dots,A_{N_p-1}^\kappa(s)\big)^\top, \\
A_i^\kappa(s)
&=
\frac{1}{\Lambda_t^\kappa(s)}
\sum_{\alpha=1}^{N_p}\lambda_\alpha^\kappa(s)\,
\frac{\partial p_{c,\alpha}^{\mathrm{AM}}(s)}{\partial s_i}.
\label{eq:A-field}
\end{align}
$\mathbf A^\kappa$ is a vector field on the saturation space (one component per independent saturation). It is \textbf{not} a potential; rather, it collects the mobility–weighted capillary gradients.  
\emph{Units.} Since $\lambda_\alpha/\Lambda_t$ is dimensionless and $\partial p_c/\partial s_i$ has units of pressure, each $A_i^\kappa$ has units of pressure.  
\emph{gTD criterion.} A scalar potential $\Pi^\kappa(s)$ exists \emph{iff} $\mathbf A^\kappa$ is curl–free on the simplex:
\begin{align}
\frac{\partial A_i^\kappa}{\partial s_j}(s)
=
\frac{\partial A_j^\kappa}{\partial s_i}(s),
\qquad i,j=1,\dots,N_p-1,
\label{eq:gTD}
\end{align}
which reduces to \eqref{eq:TD-condition} when \(N_p=3\). In that case, $\nabla_s \Pi^\kappa=\mathbf A^\kappa$ and one may recover
\(
\Pi^\kappa(s)=\int_{\gamma} \mathbf A^\kappa(s)\!\cdot\! d s
\)
with path–independent value.

With \(p_{c,\alpha}^{\kappa,\mathrm{dyn}}=p_{c,\alpha}^{\mathrm{AM}}(S)-\tau_\alpha^\kappa(S)\partial_t S_\alpha^\kappa\) from \eqref{eq:pc-dynamic}, the weighted capillary term splits as
\begin{align}
\frac{1}{\Lambda_t^\kappa}\sum_\alpha \lambda_\alpha^\kappa\,\nabla p_{c,\alpha}^{\kappa,\mathrm{dyn}}
&=
\underbrace{\frac{1}{\Lambda_t^\kappa}\sum_\alpha \lambda_\alpha^\kappa\,\nabla p_{c,\alpha}^{\mathrm{AM}}(S)}_{\text{equilibrium part}} \nonumber \\
&-
\underbrace{\frac{1}{\Lambda_t^\kappa}\sum_\alpha \lambda_\alpha^\kappa\,\nabla\!\big(\tau_\alpha^\kappa(S)\,\partial_t S_\alpha^\kappa\big)}_{\text{pseudo-parabolic part}}.
\label{eq:split-dyn}
\end{align}
TD/gTD apply to the equilibrium piece (replace \(p_{c,\alpha}^\kappa\) by \(p_{c,\alpha}^{\mathrm{AM}}\) in \eqref{eq:TD-condition}/\eqref{eq:gTD}), yielding \(\nabla \Pi^\kappa(S)\). The pseudo–parabolic part has no saturation–only potential and remains in the transport drifts; it does not affect \eqref{eq:pg} or \eqref{eq:vt-gp}.

When \eqref{eq:TD-condition} or \eqref{eq:gTD} is violated, we still \emph{define} \(p_g^\kappa\) via \eqref{eq:pg} and retain \eqref{eq:vt-gp}. As a surrogate equilibrium potential, solve the $H^1$–orthogonal projection: find \(\Phi^\kappa\in H^1(\Omega)\) with \(\int_\Omega \Phi^\kappa\,dx=0\) such that
\begin{align}
&\int_\Omega \nabla\Phi^\kappa\!\cdot\!\nabla\psi\,dx
=\int_\Omega \Big(\tfrac{1}{\Lambda_t^\kappa}\sum_{\alpha}\lambda_\alpha^\kappa\,\nabla p_{c,\alpha}^{\mathrm{AM}}\Big)\!\cdot\!\nabla\psi\,dx, \label{eq:least-squares} \\
&\forall\,\psi\in H^1(\Omega). \nonumber 
\end{align}
Then 
\(\mathbf R^\kappa:=\frac{1}{\Lambda_t^\kappa}\sum_\alpha\lambda_\alpha^\kappa\,\nabla p_{c,\alpha}^{\mathrm{AM}}-\nabla\Phi^\kappa\)
is weakly divergence–free and vanishes when TD/gTD holds. Using \(\Phi^\kappa\) recovers the exact potential wherever compatible and yields a conservative approximation otherwise; the dynamic term \eqref{eq:split-dyn} remains explicit.

With \eqref{eq:vt-gp}, the incompressible pressure problem is
\begin{align}
-\nabla\!\cdot\!\big(k^\kappa\,\Lambda_t^\kappa\,\nabla p_g^\kappa\big)
= q_t^\kappa,
\qquad q_t^\kappa:=\sum_{\alpha} q_\alpha^\kappa ,
\label{eq:pg-elliptic}
\end{align}
and becomes weakly parabolic once compressibilities appear in \eqref{eq:comp-balance-molar}.  
Forchheimer corrections do not alter the \emph{definition} \eqref{eq:pg}; operationally, we (i) solve \eqref{eq:pg-elliptic} for \(p_g^\kappa\), (ii) compute \(\mathbf v_t^{\kappa,D}\) from \eqref{eq:vt-gp}, (iii) reconstruct phase fluxes using the Darcy split \eqref{eq:darcy-split} (or the damped split \eqref{eq:operational-split}), and (iv) advance the conservative transport \eqref{eq:comp-balance-molar}.

\subsection{Classical limit and well-posedness}\label{subsec:classical-wellposed}

The purpose of this section is twofold: First, it shows that our GBL-\(N\) equations reduce exactly to the classical Buckley--Leverett (BL) model when all additional physics are switched off. Second, it sketches an energy estimate showing that Maxwell--Stefan (MS) diffusion \eqref{eq:MSz}--\eqref{eq:MSmu} together with dynamic capillarity \eqref{eq:pc-dynamic} yields a strictly (pseudo-)parabolic transport operator for \((S,z)\) and hence a well-posed mixed (elliptic--parabolic) system.\footnote{Here ``well-posed'' means: existence, uniqueness, and continuous dependence on data for the initial--boundary-value problem. In our formulation, pressure solves an elliptic (or weakly parabolic) problem, while the transport unknowns obey a pseudo-parabolic system; the purely hyperbolic BL limit is recovered only when all dissipative terms vanish.}

A fixed continuum \(\kappa\) is assumed, and \({}^\kappa\) is suppressed for readability. The standard BL hypotheses are imposed:
\emph{(a)} incompressible rock and fluids (\(\phi\) and \(\rho_\alpha\) constants);
\emph{(b)} fixed phase compositions and no interphase mass exchange (each \(x_{i\alpha}\) constant, hence each phase molar density \(c_\alpha\) constant up to \(S_\alpha\) factors);
\emph{(c)} no multicomponent diffusion/dispersion (\(\mathbf J_{i\alpha}\equiv\mathbf 0\) in \eqref{eq:MSz}--\eqref{eq:MSmu});
\emph{(d)} negligible capillarity (\(p_{c,\alpha}^{\mathrm{AM}}\equiv 0\) and \(\tau_\alpha\equiv 0\) in \eqref{eq:pc-dynamic});
\emph{(e)} Darcy momentum (set \(\beta_\alpha=0\) in \eqref{eq:momentum-nd});
\emph{(f)} constant permeability \(k\equiv k_0\) (freeze \eqref{eq:stress-laws}).
Aggregating the conservative balances \eqref{eq:comp-balance-molar} over the components that reside in phase \(\alpha\) (and defining a phase source \(q_\alpha\) consistent with the fixed-composition limit) gives the phase saturation laws
\begin{align}
&\phi\,\partial_t S_\alpha + \nabla\!\cdot \mathbf v_\alpha = q_\alpha,
\label{eq:sat-balance}
\end{align}
with $\alpha=1,\dots,N_p,\quad \sum_\alpha S_\alpha=1,$ so there are \(N_p-1\) independent conservation equations. Using the exact Darcy-level split \eqref{eq:phase-Darcy}--\eqref{eq:darcy-split} with \(p_{c,\alpha}^{\mathrm{dyn}}\equiv 0\) yields
\begin{align}
\mathbf v_\alpha
&= F_\alpha(S)\,\mathbf v_t^{D}
+ k_0\,\lambda_\alpha(S)\,\big(\rho_\alpha-\bar\rho_\lambda(S)\big)\,\mathbf g, \\
F_\alpha(S)&:=\frac{\lambda_\alpha(S)}{\Lambda_t(S)}, \qquad
\bar\rho_\lambda(S):=\frac{\sum_\beta \lambda_\beta(S)\,\rho_\beta}{\Lambda_t(S)},
\label{eq:BL-vsplit-reuse}
\end{align}
where \(\mathbf v_t^{D}\) is \eqref{eq:total-Darcy} evaluated with \(p_{c,\alpha}^{\mathrm{dyn}}=0\).
If gravity is dropped (\(\mathbf g=\mathbf 0\)), \eqref{eq:sat-balance} becomes the textbook BL system with fluxes \(F_\alpha(S)\,\mathbf v_t^{D}\); retaining gravity gives the standard buoyancy correction.

In the classical three-phase BL limit (no capillarity, no diffusion), the saturation subsystem involves two conservation laws \((S_1,S_2)\) with flux \(\boldsymbol F(S)\). The Jacobian \(\mathbf J(S)=\partial \boldsymbol F/\partial S\) has two eigenvalues \(\lambda_1(S),\lambda_2(S)\).
\emph{Strict hyperbolicity} means these eigenvalues are real and distinct.  
An \emph{umbilic point} is a state \(S^\star\) where the eigenvalues \emph{coincide}, \(\lambda_1(S^\star)=\lambda_2(S^\star)\), and the fields are not genuinely nonlinear; nearby, the characteristic structure becomes degenerate.  
An \emph{elliptic pocket} is a region of the ternary saturation diagram where the discriminant is negative and the eigenvalues become \emph{complex} (loss of hyperbolicity). In practice this causes non-uniqueness/instability of Riemann solutions and grid-dependent results unless a regularization (diffusion and/or capillarity) is present. Our GBL-\(N\) model reproduces this pathology in the classical limit and removes it once MS diffusion and/or dynamic capillarity are switched on; see below.

Activate the two regularizing mechanisms central to GBL-\(N\): (1) MS diffusion \eqref{eq:MSz}--\eqref{eq:MSmu} with symmetric positive definite (SPD) porous-media tensors, and (2) dynamic capillarity \eqref{eq:pc-dynamic} with \(\tau_\alpha(S)>0\).
We outline the dissipation they induce under periodic or no-flux boundaries.

Testing \eqref{eq:comp-balance-molar} with chemical potentials and summing over components/phases gives the standard dissipation identity
\begin{align}
&\sum_{\alpha}\int_\Omega \sum_{i} \mathbf J_{i\alpha}\!\cdot\!\nabla \mu_{i\alpha}\,dx
= \nonumber \\
&-\sum_{\alpha}\int_\Omega 
c_\alpha\,\nabla \boldsymbol{x}_\alpha^{\!\top}\,
\mathbf D_{\alpha}^{\mathrm{MS}}\,
\TF_\alpha\,
\nabla \boldsymbol{x}_\alpha\,dx
\;\le\;0,
\label{eq:MS-diss-reuse}
\end{align}
with \(\TF_\alpha=\partial \boldsymbol{\mu}_\alpha/\partial(\ln \boldsymbol{x}_\alpha)\) SPD in stable phases. Control of \(\nabla \boldsymbol{x}_\alpha\) lifts, via the EOS map \eqref{eq:state-map}--\eqref{eq:fugacity}, to control of \(\nabla z\).

Substituting \(p_{c,\alpha}^{\mathrm{dyn}}=p_{c,\alpha}^{\mathrm{AM}}(S)-\tau_\alpha(S)\,\partial_t S_\alpha\) into \eqref{eq:momentum-nd} and back into the saturation balances extracted from \eqref{eq:comp-balance-molar} yields the pseudo-parabolic form
\begin{align}
\partial_t(\phi S_\alpha)
+\nabla\!\cdot(\ldots)
-\nabla\!\cdot\!\Big(M_\alpha(S)\,\tau_\alpha(S)\,\nabla(\partial_t S_\alpha)\Big)
= \ldots,
\label{eq:pseudoparabolic-reuse}
\end{align}
with \(M_\alpha(S):=k\,\lambda_\alpha(S)\), where \((\ldots)\) denotes advective and equilibrium-capillary terms (the latter \(\propto \partial_S p_{c}^{\mathrm{AM}}\,\nabla S\)).
Testing \eqref{eq:pseudoparabolic-reuse} with \(\partial_t S_\alpha\) and summing over phases gives
\begin{align}
&\frac{d}{dt}\,\mathcal E_c(S)
+\sum_{\alpha}\int_\Omega M_\alpha(S)\,\tau_\alpha(S)\,
\big|\nabla(\partial_t S_\alpha)\big|^2\,dx \nonumber \\
&\le\; \text{advective work} + \text{MS coupling},
\label{eq:cap-energy-reuse}
\end{align}
for a capillary energy \(\mathcal E_c(S)\) associated with \(p_{c}^{\mathrm{AM}}(S)\). The coercive term controls \(\nabla(\partial_t S)\).

\emph{Energy inequality.} Combining \eqref{eq:MS-diss-reuse} and \eqref{eq:cap-energy-reuse}, and bounding advection and source terms by data, we obtain the schematic estimate
\begin{align}
\frac{d}{dt}\,\Big[\mathcal F(z) &+ \mathcal E_c(S)\Big]
+
\underbrace{\sum_{\alpha}\!\int_\Omega c_\alpha\,\nabla \boldsymbol{x}_\alpha^{\!\top}\,
\mathbf D_{\alpha}^{\mathrm{MS}}\,
\TF_\alpha\,
\nabla \boldsymbol{x}_\alpha\,dx}_{\text{controls }\nabla z} \nonumber \\
&+
\underbrace{\sum_{\alpha}\!\int_\Omega M_\alpha\,\tau_\alpha\,\big|\nabla(\partial_t S_\alpha)\big|^2\,dx}_{\text{controls }\nabla(\partial_t S)}
\;\le\; \mathcal R(t),
\label{eq:energy-ineq-reuse}
\end{align}
where \(\mathcal F(z)\) is the EOS-induced mixture free energy (convex in \(z\) under standard stability) and \(\mathcal R(t)\) depends on bounded data. Hence the transport part is strictly (pseudo-)parabolic: MS diffusion smooths compositions, and dynamic capillarity regularizes saturations in the Hassanizadeh--Gray sense. The pressure subproblem remains elliptic (or weakly parabolic) via \eqref{eq:vt-gp}. In axisymmetry, the same estimates hold with \(\nabla,\nabla\!\cdot\) replaced by their cylindrical forms (see §\ref{subsec:axisym}).

If \(\mathbf D_{\alpha}^{\mathrm{MS}}\!\to\!\mathbf 0\) and \(\tau_\alpha\!\to\!0\), the dissipation in \eqref{eq:energy-ineq-reuse} collapses and the system reverts to the BL hyperbolic limit, with the well-known loss of strict hyperbolicity for \(N_p\!\ge\!3\) (umbilic points and elliptic pockets as explained above). Any strictly positive MS diffusion and/or dynamic capillarity provides a priori bounds independent of grid size, which is precisely how the GBL-\(N\) model attains well-posedness in regimes where classical BL fails.

\subsection{Fracture modeling and non-Darcy effects}\label{subsec:fractures}

We use the same strictly conservative balances \eqref{eq:comp-balance-molar} in both continua—matrix ($\kappa=\mathrm m$) and fractures ($\kappa=\mathrm f$); differences enter only through geometry and constitutive data. In fractures we work with \emph{aperture-integrated} (areal) fluxes defined on the fracture plane (units m$^2$/s). For a lower-dimensional fracture control volume, the divergence is the \emph{tangential} (surface) divergence $\nabla_\tau\!\cdot$ acting on areal fluxes.

(i) \emph{Dual–continuum}: matrix–fracture coupling appears as conservative sources $T_i^{\mathrm m\leftrightarrow \mathrm f}$ assembled via \eqref{eq:WR-transfer}–\eqref{eq:WR-component}. 
(ii) \emph{EDFM/pEDFM}: fractures are lower-dimensional control volumes coupled to matrix cells by NNCs using \eqref{eq:NNC-phase}–\eqref{eq:NNC-component} \citep{WarrenRoot1963,Olorode2020pEDFM}.

Let $b$ be the hydraulic aperture and $\zeta_r\!\in\!(0,1]$ a conductivity reduction factor: $\zeta_r=1$ gives the smooth parallel-plate cubic law, while smaller values represent roughness/contact-induced reductions; $\zeta_r\to 0$ corresponds to an effectively closed/nonconductive pathway.
The in-plane permeability and transmissivity are
\begin{align}
k^{\mathrm f}&=\frac{\zeta_r\, b^2}{12}, \\
\mathcal T^{\mathrm f}&=k^{\mathrm f}\, b=\frac{\zeta_r\, b^3}{12},
\label{eq:fracture-kT}
\end{align}
and a stress–aperture law consistent with \eqref{eq:stress-laws} is
\begin{align}
b(\sigma')&=b_0\,e^{-\gamma_b(\sigma'-\sigma_0')}, \\
k^{\mathrm f}(\sigma')&=\frac{\zeta_r\, b(\sigma')^2}{12}, \\
\mathcal T^{\mathrm f}(\sigma')&=\frac{\zeta_r\, b(\sigma')^3}{12}.
\label{eq:fracture-stress}
\end{align}

Specializing \eqref{eq:phase-Darcy} to tangential gradients $\nabla_\tau=(\mathbf I-\mathbf n\otimes\mathbf n)\nabla$ gives the aperture-integrated (areal) phase flux
\begin{align}
\mathbf v_{\alpha}^{\mathrm f,D}
&=
-\,\mathcal T^{\mathrm f}\,\lambda_\alpha^{\mathrm f}(S^{\mathrm f})\,
\Big(\nabla_\tau p^{\mathrm f}-\rho_\alpha^{\mathrm f}\,\mathbf g_\tau-\nabla_\tau p_{c,\alpha}^{\mathrm f,\mathrm{dyn}}\Big), \\
\mathbf g_\tau&=(\mathbf I-\mathbf n\otimes\mathbf n)\mathbf g,
\label{eq:fracture-darcy}
\end{align}
with $p_{c,\alpha}^{\mathrm f,\mathrm{dyn}}$ from \eqref{eq:pc-dynamic} using fracture parameters $(\gamma,\theta,\mathcal M_{\mathrm f},\xi_{\mathrm f})$. Densities $\rho_\alpha^{\mathrm f}$ need not be constant; the body-force form is consistent in compressible/compositional settings.

Inertial deviations are strongest in fractures. We therefore take $\beta_\alpha^{\mathrm m}=0$ in the matrix (unless core data show otherwise) and retain $\beta_\alpha^{\mathrm f}>0$ in fractures. The same per-phase damping defined in \eqref{eq:chi-factor}–\eqref{eq:chi-linear} is used here with fracture properties:
\begin{align}
\mathbf v_\alpha^{\mathrm f}=\chi_\alpha^{\mathrm f}\,\mathbf v_{\alpha}^{\mathrm f,D}, 
\qquad
u_{\alpha}^{\mathrm f,D}:=\|\mathbf v_{\alpha}^{\mathrm f,D}\|,
\label{eq:fracture-forchheimer}
\end{align}
where $\chi_\alpha^{\mathrm f}$ is given by \eqref{eq:chi-factor} after the substitutions 
$\{k,k_{r\alpha},\mu_\alpha,\rho_\alpha,u_{\alpha}^{D}\}\to\{k^{\mathrm f},k_{r\alpha}^{\mathrm f},\mu_\alpha^{\mathrm f},\rho_\alpha^{\mathrm f},u_{\alpha}^{\mathrm f,D}\}$.
For weak inertia, use the linearization \eqref{eq:chi-linear}. 
A convenient activation metric is the in-plane Reynolds number
\(
\mathrm{Re}_\tau^\alpha=(\rho_\alpha^{\mathrm f}\|\mathbf v_{\alpha}^{\mathrm f,D}\|\,b)/\mu_\alpha^{\mathrm f}.
\)
Rough-walled fracture studies report measurable departures from Darcy/cubic-law behavior once $\mathrm{Re}_\tau^\alpha=\mathcal O(1)$, while Forchheimer-type quadratic corrections often provide good fits in the stronger-inertia regime $\mathrm{Re}_\tau^\alpha\gtrsim 20$ (geometry/roughness dependent) \citep{AlYaarubi2003,Yin2018Processes}. Consistent with these observations, increasing roughness and decreasing hydraulic aperture tend to enhance inertial losses; accordingly, the Forchheimer coefficient is not universal in fractures and is commonly modeled as $\beta_\alpha^{\mathrm f}=\beta_\alpha^{\mathrm f}(b,\mathrm{roughness})$ with $b=b(\sigma')$, with $\beta_\alpha^{\mathrm f}$ typically increasing as aperture decreases (often approximated by inverse-aperture trends over limited ranges) \citep{Chen2015JHydrol}.

To remain strictly conservative while accounting for high-rate inertia at interfaces:
\begin{itemize}
\item Dual–continuum transfer \eqref{eq:WR-transfer}–\eqref{eq:WR-component}: scale each phase transfer by the fracture-side damping, 
$\ \mathcal T_{\alpha,\mathrm{ND}}^{\mathrm m\leftrightarrow \mathrm f}=\chi_\alpha^{\mathrm f}\,\mathcal T_\alpha^{\mathrm m\leftrightarrow \mathrm f}.$
\item EDFM/pEDFM NNCs \eqref{eq:NNC-phase}–\eqref{eq:NNC-component}: scale the phase flux,
$\ F_{\alpha,\mathrm{ND}}^{m\to f}=\chi_\alpha^{\mathrm f}\,F_{\alpha}^{m\to f}.$
\end{itemize}
The same factor multiplies equal-and-opposite interface fluxes, so conservation is unchanged.

The definition \eqref{eq:pg} applies tangentially:
\begin{align}
\sum_{\alpha}\lambda_\alpha^{\mathrm f}(S^{\mathrm f})\,
\big(\nabla_\tau p_\alpha^{\mathrm f}-\rho_\alpha^{\mathrm f}\mathbf g_\tau\big)
&= \Lambda_t^{\mathrm f}(S^{\mathrm f})\,\nabla_\tau p_g^{\mathrm f}, \\
\mathbf v_t^{\mathrm f,D}&=-\,\mathcal T^{\mathrm f}\,\Lambda_t^{\mathrm f}\,\nabla_\tau p_g^{\mathrm f},
\label{eq:pg-fracture}
\end{align}
the fracture analogues of \eqref{eq:pg} and \eqref{eq:vt-gp}. TD/gTD from \S\ref{subsec:global-pressure} carries over unchanged, now on the fracture plane. 
(For radial/axisymmetric problems in the matrix, use the cylindrical operators in \S\ref{subsec:axisym}; the fracture-plane equations remain tangential.)

\medskip
Fractures are therefore modeled within the same conservative framework: geometry enters via \eqref{eq:fracture-kT}–\eqref{eq:fracture-stress}; Darcy tangential flow follows \eqref{eq:fracture-darcy}; non-Darcy acts through the already-defined damping \eqref{eq:chi-factor}–\eqref{eq:chi-linear} as in \eqref{eq:fracture-forchheimer}; and coupling terms are scaled without breaking conservation. The global-pressure machinery remains valid in fractures via \eqref{eq:pg-fracture}.

\subsection{Global Buckley--Leverett-$N$: operative system}\label{subsec:gbln-operative}

In each continuum $\kappa\in\{\mathrm m,\mathrm f\}$ we advance the global pressure $p_g^\kappa$, the phase saturations $\{S_\alpha^\kappa\}_{\alpha=1}^{N_p}$ with $\sum_\alpha S_\alpha^\kappa=1$, and the overall composition $z=(z_1,\dots,z_{N_c})$ with $\sum_i z_i=1$.
Phase/PVT data $\{\nu_\alpha,x_{i\alpha},\rho_\alpha,\mu_\alpha\}$ are obtained from the isothermal EOS flash \eqref{eq:state-map}--\eqref{eq:fugacity}. 
Mobilities are $\lambda_\alpha^\kappa(S)=k_{r\alpha}^\kappa(S)/\mu_\alpha^\kappa$; intrinsic permeability $k^\kappa(p,\sigma')$ follows \eqref{eq:stress-laws}. 
Dynamic capillarity uses \eqref{eq:pc-dynamic} with $p_{c,\alpha}^{\mathrm{AM}}(S)$ from \eqref{eq:AM-pcS}. 
The (post-damping) \emph{true} total superficial flux is $\mathbf v_t^\kappa:=\sum_{\beta=1}^{N_p}\mathbf v_\beta^\kappa$.

\medskip

With per-phase Forchheimer damping $\chi_\alpha^\kappa$ from \eqref{eq:chi-factor} (or \eqref{eq:chi-linear}), define the apparent mobilities/weights
\begin{align}
\tilde{\lambda}_\alpha^\kappa &:= \chi_\alpha^\kappa\,\lambda_\alpha^\kappa, \\
\tilde{\Lambda}_t^\kappa &:= \sum_{\beta=1}^{N_p}\tilde{\lambda}_\beta^\kappa, \\
\tilde F_\alpha^\kappa &:= \frac{\tilde{\lambda}_\alpha^\kappa}{\tilde{\Lambda}_t^\kappa},
\label{eq:GBL-apparent}
\end{align}
and the corresponding mobility-weighted averages
\begin{align}
\tilde{\bar\rho}_\lambda^\kappa
&:=\frac{1}{\tilde{\Lambda}_t^\kappa}\sum_{\beta=1}^{N_p}\tilde{\lambda}_\beta^\kappa\,\rho_\beta^\kappa, \\
\widetilde{\overline{\nabla p_c}}^{\,\kappa}
&:=\frac{1}{\tilde{\Lambda}_t^\kappa}\sum_{\beta=1}^{N_p}\tilde{\lambda}_\beta^\kappa\,\nabla p_{c,\beta}^{\kappa,\mathrm{dyn}}.
\label{eq:GBL-averages}
\end{align}
Then each phase flux is advanced by
\begin{align}
\mathbf v_\alpha^\kappa
=
\tilde F_\alpha^\kappa\,\mathbf v_t^\kappa
&+
k^\kappa\,\tilde{\lambda}_\alpha^\kappa
\Big[
\big(\rho_\alpha^\kappa-\tilde{\bar\rho}_\lambda^\kappa\big)\,\mathbf g \nonumber \\
&\qquad+
\big(\nabla p_{c,\alpha}^{\kappa,\mathrm{dyn}}-\widetilde{\overline{\nabla p_c}}^{\,\kappa}\big)
\Big],
\label{eq:GBL-master}
\end{align}
the operational (Forchheimer-aware) version of the exact Darcy split \eqref{eq:darcy-split}. 
\emph{Axisymmetry:} in cylindrical coordinates, replace $\nabla$ and $\nabla\!\cdot$ by their axisymmetric forms as in \S\ref{subsec:axisym}.

\medskip

The global pressure is defined by the mobility-weighted phase relation \eqref{eq:pg}. 
Consequently,
\begin{align}
&\mathbf v_t^{\kappa,D} = -\,k^\kappa\,\Lambda_t^\kappa\,\nabla p_g^\kappa, 
\qquad \Lambda_t^\kappa=\sum_{\beta=1}^{N_p}\lambda_\beta^\kappa,
\label{eq:vt-gp-reuse}
\\[-0.25em]
&-\nabla\!\cdot\!\big(k^\kappa\,\Lambda_t^\kappa\,\nabla p_g^\kappa\big) = q_t^\kappa, 
\qquad q_t^\kappa=\sum_{\alpha=1}^{N_p}q_\alpha^\kappa,
\label{eq:pg-elliptic-reuse}
\end{align}
i.e.\ \eqref{eq:vt-gp}–\eqref{eq:pg-elliptic}. 
Solve \eqref{eq:pg-elliptic-reuse} for $p_g^\kappa$ (Cartesian or axisymmetric form per \S\ref{subsec:axisym}), obtain $\mathbf v_t^{\kappa,D}$ from \eqref{eq:vt-gp-reuse}, and reconstruct the phase fluxes via \eqref{eq:GBL-master}. 
(Generally $\mathbf v_t^\kappa\neq \mathbf v_t^{\kappa,D}$ because each phase has its own damping $\chi_\alpha^\kappa$.)

\medskip

The EOS flash needs a scalar \emph{thermodynamic} pressure $p^\kappa$. Construct it from $p_g^\kappa$ without embedding gravity (to avoid spurious $\nabla\rho$ terms):
\begin{itemize}
\item If TD/gTD holds (\S\ref{subsec:global-pressure}), set 
$
p^\kappa := p_g^\kappa + \Pi^\kappa(S) + C^\kappa,
$ 
where $\Pi^\kappa$ is the saturation potential and $C^\kappa$ a constant chosen to honor a pressure datum (well/Dirichlet or mean).
\item If TD/gTD fails, use the conservative surrogate potential from the $H^1$ projection \eqref{eq:least-squares} and set 
$
p^\kappa := p_g^\kappa + \Phi^\kappa + C^\kappa.
$
\end{itemize}
This is consistent with the body-force form in \eqref{eq:pg} and with the fractional-flow split, while providing the $p^\kappa$ required by the EOS flash \eqref{eq:state-map}–\eqref{eq:fugacity}.

\medskip

For each component $i=1,\dots,N_c$,
\begin{align}
&\frac{\partial}{\partial t}
\Big[
\phi^\kappa \sum_{\alpha=1}^{N_p} S_\alpha^\kappa\, c_\alpha^\kappa\, x_{i\alpha}^\kappa
+\rho_r^\kappa\, \Gamma_i^\kappa
\Big] \nonumber \\
&\quad+
\nabla\!\cdot
\Big[
\sum_{\alpha=1}^{N_p} c_\alpha^\kappa x_{i\alpha}^\kappa\,\mathbf v_\alpha^\kappa
-\sum_{\alpha=1}^{N_p}\mathbf J_{i\alpha}^\kappa
\Big]
=
q_i^\kappa + T_i^{\mathrm{m}\leftrightarrow \mathrm{f}} .
\label{eq:GBL-transport}
\end{align}
Here $\mathbf J_{i\alpha}^\kappa$ are Maxwell–Stefan fluxes \eqref{eq:MSz}–\eqref{eq:MSmu} with the constraint \eqref{eq:MSconstraint}; adsorption follows \eqref{eq:ads-kinetics}; and matrix–fracture exchange is given either by \eqref{eq:WR-transfer}–\eqref{eq:WR-component} (dual–continuum) or by \eqref{eq:NNC-phase}–\eqref{eq:NNC-component} (EDFM/pEDFM). 
Setting $\chi_\alpha^\kappa=1$, $\tau_\alpha^\kappa=0$, $\mathbf J_{i\alpha}^\kappa=\mathbf 0$, and freezing $k^\kappa$ reduces \eqref{eq:GBL-master}–\eqref{eq:GBL-transport} to classical BL with $F_\alpha=\lambda_\alpha/\Lambda_t$ (see \S\ref{subsec:classical-wellposed}). 
For fractures, reuse the same structure with fracture geometry/stress encoded in $k^\kappa$ and $\chi_\alpha^\kappa$ (\S\ref{subsec:fractures}). 
When TD/gTD holds, the equilibrium capillary drift admits a scalar potential via \eqref{eq:TD-condition}–\eqref{eq:gTD}; otherwise employ the projection \eqref{eq:least-squares}.

\medskip

Given $(S^n,z^n,p_g^{\kappa,\ell})$:
\begin{enumerate}
\item \emph{EOS flash:} using a provisional $p^{\kappa,\ell}$ (from $p_g^{\kappa,\ell}$ and $\Pi^\kappa$ or $\Phi^\kappa$), compute $\{\nu_\alpha,x_{i\alpha},\rho_\alpha,\mu_\alpha\}$; update $\lambda_\alpha^\kappa$, $p_{c,\alpha}^{\kappa,\mathrm{dyn}}$, and $\mathbf D^{\mathrm{MS}}$.
\item \emph{Pressure:} assemble $\Lambda_t^\kappa$ and $k^\kappa$; solve \eqref{eq:pg-elliptic-reuse} for $p_g^{\kappa,\ell+1}$ (Cartesian or axisymmetric).
\item \emph{Fluxes:} compute $\mathbf v_t^{\kappa,D}$ via \eqref{eq:vt-gp-reuse}; evaluate $\chi_\alpha^\kappa$ and form $\mathbf v_\alpha^\kappa$ from \eqref{eq:GBL-master}.
\item \emph{Transport:} advance \eqref{eq:GBL-transport} for $(S,z)$ with MS diffusion \eqref{eq:MSz}–\eqref{eq:MSmu} and dynamic capillarity \eqref{eq:pc-dynamic}; include $T_i^{\mathrm m\leftrightarrow \mathrm f}$.
\item \emph{Update EOS pressure:} set $p^{\kappa,\ell+1}=p_g^{\kappa,\ell+1}+\Pi^\kappa$ (or $+\Phi^\kappa$) up to a datum; re-flash if using a fully coupled iterate.
\end{enumerate}
Steps 1–5 are iterated to a chosen nonlinearity tolerance (sequential or monolithic).

\subsection{Axisymmetric (cylindrical) form for radial injection with vertical buoyancy}
\label{subsec:axisym}

We impose axisymmetry in $(r,z)$ with no $\theta$-dependence and gravity $\mathbf g=-\,g\,\hat{\boldsymbol z}$. 
All expressions below are the cylindrical rewrite of the global BL system in \S\ref{subsec:gbln-operative}, using the same closures and references.

For a vector $\mathbf a=(a_r,a_z)$ and a scalar $\phi$,
\begin{align}
\nabla\!\cdot \mathbf a &= \frac{1}{r}\frac{\partial}{\partial r}\!\big(r a_r\big)+\frac{\partial a_z}{\partial z}, \\
\nabla \phi &= \Big(\frac{\partial \phi}{\partial r},\,\frac{\partial \phi}{\partial z}\Big).
\label{eq:cyl-ops}
\end{align}

With the definition of global pressure \eqref{eq:pg}, the pressure equation \eqref{eq:pg-elliptic} becomes
\begin{align}
-\,\nabla\!\cdot\!\big(k^\kappa \Lambda_t^\kappa \nabla p_g^\kappa\big)
= q_t^\kappa
\ \Longleftrightarrow\ 
&-\frac{1}{r}\frac{\partial}{\partial r}
\Big(r\,k^\kappa \Lambda_t^\kappa \frac{\partial p_g^\kappa}{\partial r}\Big) \nonumber \\
&-\frac{\partial}{\partial z}
\Big(k^\kappa \Lambda_t^\kappa \frac{\partial p_g^\kappa}{\partial z}\Big)
= q_t^\kappa.
\label{eq:cyl-pg}
\end{align}
The Darcy total-flux components from \eqref{eq:vt-gp} are
\begin{align}
v_{t,r}^{\kappa,D}&=-\,k^\kappa \Lambda_t^\kappa \frac{\partial p_g^\kappa}{\partial r}, \\
v_{t,z}^{\kappa,D}&=-\,k^\kappa \Lambda_t^\kappa \frac{\partial p_g^\kappa}{\partial z}.
\label{eq:cyl-vt}
\end{align}
(Body forces are absorbed in \eqref{eq:pg}, so \eqref{eq:cyl-pg} retains the Cartesian structure.)

Using the operational split \eqref{eq:GBL-master}, the radial and vertical components are
\begin{align}
v_{\alpha,r}^\kappa
&=\tilde F_\alpha^\kappa\,v_{t,r}^\kappa
+ k^\kappa \tilde\lambda_\alpha^\kappa
\Big[
\underbrace{\big(\rho_\alpha^\kappa-\tilde{\bar\rho}_\lambda^\kappa\big)\,\mathbf g\!\cdot\!\hat{\mathbf r}}_{=\,0} \nonumber \\
&\qquad+\Big(\partial_r p_{c,\alpha}^{\kappa,\mathrm{dyn}}
-\widetilde{\overline{\partial_r p_c}}^{\,\kappa}\Big)
\Big],
\label{eq:cyl-vr}
\\
v_{\alpha,z}^\kappa
&=\tilde F_\alpha^\kappa\,v_{t,z}^\kappa
+ k^\kappa \tilde\lambda_\alpha^\kappa
\Big[
-\big(\rho_\alpha^\kappa-\tilde{\bar\rho}_\lambda^\kappa\big)\,g \nonumber \\
&\qquad+\Big(\partial_z p_{c,\alpha}^{\kappa,\mathrm{dyn}}
-\widetilde{\overline{\partial_z p_c}}^{\,\kappa}\Big)
\Big],
\label{eq:cyl-vz}
\end{align}
where $\tilde F_\alpha^\kappa$, $\tilde\lambda_\alpha^\kappa$, $\tilde{\bar\rho}_\lambda^\kappa$ and the capillary averages are as in \eqref{eq:GBL-apparent}–\eqref{eq:GBL-averages}, and $p_{c,\alpha}^{\kappa,\mathrm{dyn}}$ is given by \eqref{eq:pc-dynamic}. 
Note that $\mathbf v_t^\kappa=\sum_\alpha \mathbf v_\alpha^\kappa$ is the \emph{true} total flux (generally different from $\mathbf v_t^{\kappa,D}$ when Forchheimer damping is active). 
Buoyancy has no radial component but drives vertical segregation through \eqref{eq:cyl-vz}.

From \eqref{eq:GBL-transport} and \eqref{eq:cyl-ops}, for each component $i=1,\dots,N_c$,
\begin{align}
&\frac{\partial}{\partial t}
\Big[
\phi^\kappa \sum_{\alpha=1}^{N_p} S_\alpha^\kappa c_\alpha^\kappa x_{i\alpha}^\kappa
+ \rho_r^\kappa \Gamma_i^\kappa
\Big]
+\frac{1}{r}\frac{\partial}{\partial r}
\Big(
r \sum_{\alpha=1}^{N_p} c_\alpha^\kappa x_{i\alpha}^\kappa v_{\alpha,r}^\kappa
\Big)\nonumber \\
&\quad+\frac{\partial}{\partial z}
\Big(
\sum_{\alpha=1}^{N_p} c_\alpha^\kappa x_{i\alpha}^\kappa v_{\alpha,z}^\kappa
\Big) 
-\frac{1}{r}\frac{\partial}{\partial r}
\Big(r \sum_{\alpha=1}^{N_p} (J_{i\alpha}^\kappa)_r\Big) \nonumber \\
&\quad-\frac{\partial}{\partial z}
\Big(\sum_{\alpha=1}^{N_p} (J_{i\alpha}^\kappa)_z\Big)
= q_i^\kappa+T_i^{\mathrm m\leftrightarrow \mathrm f}.
\label{eq:cyl-transport}
\end{align}
Maxwell--Stefan diffusion $\mathbf J_{i\alpha}^\kappa$ is given by \eqref{eq:MSz}--\eqref{eq:MSmu} with the constraint \eqref{eq:MSconstraint}. 
For purely radial injection one may consider an initially $z$-uniform state ($\partial_z(\cdot)=0$); vertical buoyancy then emerges through \eqref{eq:cyl-vz} as segregation develops. 
Regularity at the axis requires $r\,v_{t,r}^{\kappa,D}$ bounded as $r\to 0$ (equivalently, $\partial_r p_g^\kappa$ bounded there).

\section*{Conclusions}

We assembled a \emph{global Buckley--Leverett} (GBL--$N$) formulation that preserves the BL intuition—an explicit fractional–flow split and a scalar global pressure—while accommodating the physics required by fractured, multicomponent systems. The backbone is: the operative phase–flux split \emph{(Global BL master equation)} \eqref{eq:GBL-master}; the global–pressure definition and Darcy total–flux relation \eqref{eq:pg}, \eqref{eq:vt-gp}; and the strictly conservative multicomponent transport \eqref{eq:GBL-transport}. These are closed with EOS–consistent phase behavior via the flash map \eqref{eq:state-map}--\eqref{eq:fugacity}, ensuring thermodynamically consistent compositions, densities, and viscosities.

The formulation addresses the loss of strict hyperbolicity in multi–phase BL with two minimal, physics–anchored regularizations. First, Maxwell--Stefan diffusion \eqref{eq:MSz}, together with the no–net–diffusion constraint \eqref{eq:MSconstraint} and the chemical–potential form \eqref{eq:MSmu}, supplies an SPD dissipation on composition gradients. Second, dynamic capillarity \eqref{eq:pc-dynamic}, built on the morphology–based equilibrium law \eqref{eq:AM-pcS}, enters the split \eqref{eq:GBL-master} and produces both capillary smoothing and the pseudo–parabolic contribution made explicit in \eqref{eq:momentum-expanded}. Their combined effect is quantified by the energy inequality in \S\ref{subsec:classical-wellposed} (see \eqref{eq:energy-ineq-reuse}): MS diffusion damps $\nabla z$ and dynamic $p_c$ controls $\nabla(\partial_t S)$, restoring well–posedness away from the classical BL limit.

On pressure–transport decoupling, the mobility–weighted global pressure \eqref{eq:pg} always exists and yields the elliptic/weakly–parabolic total–flux form \eqref{eq:vt-gp}. When TD (or gTD) compatibility holds, cf.\ \eqref{eq:TD-condition}, \eqref{eq:gTD}, the equilibrium capillary drift collapses to a saturation potential, giving full equivalence to the original equations; otherwise the $H^1$ projection \eqref{eq:least-squares} provides a conservative surrogate that becomes exact whenever compatibility is met.

Inertial departures from Darcy are embedded where they matter most—primarily in fractures—through the per–phase Forchheimer damping $\chi_\alpha^\kappa$ \eqref{eq:chi-factor} (or its weak–inertia linearization \eqref{eq:chi-linear}). These enter the \emph{apparent} mobilities and weights \eqref{eq:GBL-apparent}, \eqref{eq:GBL-averages}, preserving the BL form while reducing flux magnitudes at high rates. Geometry and stress sensitivity enter transparently via $k^\kappa(p,\sigma')$ \eqref{eq:stress-laws}, and matrix–fracture exchange remains strictly conservative under both dual–continuum transfer \eqref{eq:WR-transfer}, \eqref{eq:WR-component} and EDFM/pEDFM NNCs \eqref{eq:NNC-phase}, \eqref{eq:NNC-component} (see \S\ref{subsec:fractures}).

For coordinate–specific analyses (e.g., radial injection with vertical buoyancy relevant to CO$_2$ storage), the entire GBL system is written in cylindrical form in \S\ref{subsec:axisym}; it is a direct rewrite of \eqref{eq:GBL-master}, \eqref{eq:pg}, \eqref{eq:vt-gp}, and \eqref{eq:GBL-transport} with axisymmetric operators.

Finally, the formulation collapses to classical BL when its assumptions hold: setting $\chi_\alpha^\kappa\!\to\!1$, $\tau_\alpha^\kappa\!\to\!0$, $\mathbf J_{i\alpha}^\kappa\!\to\!\mathbf 0$, and freezing $k^\kappa$ recovers $w_\alpha=\lambda_\alpha/\Lambda_t$ (cf.\ \eqref{eq:BL-vsplit-reuse}) and the standard split (see \S\ref{subsec:classical-wellposed} and \eqref{eq:BL-vsplit-reuse}).

In short, GBL--$N$ is a single, conservative, and interpretable backbone: EOS–consistent, fracture–aware, compatible with a scalar global pressure, and provably regular once diffusion and dynamic capillarity are admitted—yet it reduces exactly to BL when appropriate. This makes it a practical foundation for discretization, calibration, and validation in the multiphase, multicomponent regimes of current interest.

%========================
\section*{Notation and acronyms}
\label{sec:notation}

\subsection*{Acronyms}
\begin{description}
\item[BL] Buckley--Leverett.
\item[EOS] Equation of state.
\item[PVT] Pressure--volume--temperature properties.
\item[GBL-$N$] Global Buckley--Leverett for $N_c$ components and $N_p$ phases.
\item[MS] Maxwell--Stefan (multicomponent diffusion).
\item[TD/gTD] Total-differential / generalized total-differential compatibility condition for global pressure.
\item[EDFM] Embedded discrete fracture model.
\item[pEDFM] Projection-based embedded discrete fracture model.
\item[NNC] Non-neighbor connection (matrix--fracture coupling).
\item[SRK] Soave--Redlich--Kwong cubic EOS.
\item[PR] Peng--Robinson cubic EOS.
\item[CDF] Cumulative distribution function.
\end{description}

\subsection*{Symbols (primary fields, thermodynamics, and transport)}
\begin{description}
\item[$\kappa\in\{\mathrm m,\mathrm f\}$] Continuum index: matrix ($\mathrm m$) and fracture ($\mathrm f$).
\item[$\alpha=1,\dots,N_p$] Phase index; $N_p^\star\le N_p$ phases may be present at equilibrium.
\item[$i=1,\dots,N_c$] Component index.
\item[$\mathbf x$] Spatial position vector.
\item[$t$] Time.
\item[$T$] Temperature (fixed; isothermal).
\item[$p^\kappa$] Thermodynamic pressure used in EOS flash (per continuum).
\item[$p_g^\kappa$] Global pressure (mobility-weighted scalar driving Darcy total flux in a continuum).
\item[$S_\alpha^\kappa$] Phase saturation in continuum $\kappa$, with $\sum_\alpha S_\alpha^\kappa=1$.
\item[$z_i$] Overall (bulk) mole fraction, with $\sum_i z_i=1$.
\item[$\nu_\alpha$] Phase molar fraction from EOS flash, with $\sum_\alpha \nu_\alpha=1$.
\item[$x_{i\alpha}$] Mole fraction of component $i$ in phase $\alpha$, with $\sum_i x_{i\alpha}=1$.
\item[$M_i$] Molar mass of component $i$.
\item[$\bar M_\alpha$] Phase mixture molar mass, $\bar M_\alpha=\sum_i M_i x_{i\alpha}$.
\item[$c_\alpha$] Phase molar density.
\item[$\rho_\alpha$] Phase mass density, $\rho_\alpha=\bar M_\alpha c_\alpha$.
\item[$\mu_\alpha$] Phase dynamic viscosity (from transport-property correlations evaluated at flashed state).
\item[$\mu_{i\alpha}$] Chemical potential of component $i$ in phase $\alpha$.
\item[$f_i^\alpha$] Fugacity of component $i$ in phase $\alpha$; $\varphi_{i\alpha}$ is the fugacity coefficient.
\end{description}

\subsection*{Fluxes, mobilities, and averaging conventions}
\begin{description}
\item[$\mathbf v_\alpha^\kappa$] Phase superficial (Darcy) velocity in continuum $\kappa$ (volumetric flux per bulk cross-sectional area).
\item[$\mathbf v_t^\kappa$] Total superficial velocity, $\mathbf v_t^\kappa=\sum_\alpha \mathbf v_\alpha^\kappa$.
\item[$\mathbf v_\alpha^{\kappa,D}$] Darcy-level (no inertia) phase velocity.
\item[$\mathbf v_t^{\kappa,D}$] Darcy-level total velocity.
\item[$k^\kappa$] Intrinsic permeability (stress sensitive).
\item[$k_{r\alpha}^\kappa(S)$] Relative permeability of phase $\alpha$.
\item[$\lambda_\alpha^\kappa$] Phase mobility, $\lambda_\alpha^\kappa=k_{r\alpha}^\kappa/\mu_\alpha^\kappa$.
\item[$\Lambda_t^\kappa$] Total mobility, $\Lambda_t^\kappa=\sum_\alpha \lambda_\alpha^\kappa$.
\item[$F_\alpha^\kappa$] Fractional-flow weight, $F_\alpha^\kappa=\lambda_\alpha^\kappa/\Lambda_t^\kappa$.
\item[$\chi_\alpha^\kappa$] Forchheimer damping factor (per phase), $0<\chi_\alpha^\kappa\le 1$.
\item[$\tilde{\lambda}_\alpha^\kappa$] Apparent (damped) mobility, $\tilde{\lambda}_\alpha^\kappa=\chi_\alpha^\kappa\lambda_\alpha^\kappa$.
\item[$\tilde{F}_\alpha^\kappa$] Apparent fractional-flow weight, $\tilde{F}_\alpha^\kappa=\tilde{\lambda}_\alpha^\kappa/\tilde{\Lambda}_t^\kappa$.
\item[$\bar{(\cdot)}_\lambda$] Mobility-weighted average using $\lambda_\alpha$ (e.g., $\bar\rho_\lambda=\sum_\alpha \lambda_\alpha\rho_\alpha/\Lambda_t$).
\item[$\tilde{(\cdot)}$] Quantity formed with apparent mobilities $\tilde{\lambda}_\alpha$ (e.g., $\tilde{\bar\rho}_\lambda$).
\end{description}

\subsection*{Diffusion, capillarity, and adsorption}
\begin{description}
\item[$\mathbf J_{i\alpha}$] Maxwell--Stefan diffusive molar flux of component $i$ in phase $\alpha$ (relative to phase-average velocity), satisfying $\sum_i\mathbf J_{i\alpha}=\mathbf 0$.
\item[$\mathbf D^{\mathrm{MS}}_{ij,\alpha}$] Effective Maxwell--Stefan diffusivity (isotropic scalar times $\mathbf I$ or anisotropic spatial tensor).
\item[$\mathbf D^{\mathrm{MS}}_\alpha$] Compact operator form of MS diffusivity for phase $\alpha$.
\item[$\TF_\alpha$] Thermodynamic-factor matrix in MS diffusion,
$[\TF_\alpha]_{ij}=\partial\mu_{i\alpha}/\partial\ln x_{j\alpha}$.
\item[$p_{c,\alpha}^{\kappa,\mathrm{dyn}}$] Dynamic capillary pressure (per phase, per continuum).
\item[$p_{c,\alpha}^{\mathrm{AM}}(S)$] Equilibrium capillary pressure from the pore-morphology (Alonso--Marroqu{\'\i}n) model or a monotone surrogate fit.
\item[$\tau_\alpha^\kappa(S)$] Dynamic capillarity coefficient (units Pa$\cdot$s).
\item[$\Gamma_i^\kappa$] Adsorbed inventory (storage term) of component $i$ in continuum $\kappa$.
\item[$t_i^\kappa$] Adsorption relaxation time (units s).
\end{description}

\subsection*{Fractures and matrix--fracture transfer}
\begin{description}
\item[$b$] Hydraulic aperture of a fracture.
\item[$\zeta_r$] Conductivity reduction factor relative to smooth parallel-plate cubic law, $\zeta_r\in(0,1]$.
\item[$k^{\mathrm f}$] Fracture in-plane permeability, $k^{\mathrm f}=\zeta_r b^2/12$.
\item[$\mathcal T^{\mathrm f}$] Fracture transmissivity, $\mathcal T^{\mathrm f}=\zeta_r b^3/12$.
\item[$\beta_\alpha^\kappa$] Forchheimer coefficient (calibrated; fracture values may depend on $b$ and roughness).
\item[$\mathrm{Re}_\tau^\alpha$] Fracture-plane Reynolds number, $\mathrm{Re}_\tau^\alpha=(\rho_\alpha^{\mathrm f}\|\mathbf v_{\alpha}^{\mathrm f,D}\|\,b)/\mu_\alpha^{\mathrm f}$.
\item[$\Phi_\alpha^\kappa$] Phase hydraulic potential,
$\Phi_\alpha^\kappa=p^\kappa-p_{c,\alpha}^{\kappa,\mathrm{dyn}}-\rho_\alpha^\kappa\mathbf g\!\cdot\!\mathbf x$.
\item[$\omega$] Dual-porosity shape factor (units 1/m$^2$).
\item[$k_{\mathrm{int}}$] Interporosity (matrix--fracture) permeability scale.
\item[$T_{mf}$] Geometric transmissibility for matrix--fracture NNC.
\item[$\nabla_\tau$] Tangential (fracture-plane) gradient operator.
\end{description}
%========================

\begin{acknowledgments}
Ch.T. would like to acknowledge the support provided by the Deanship of Research (DOR) at King Fahd University of Petroleum \& Minerals (KFUPM) for funding this work through project No. EC251017.
\end{acknowledgments}

\bibliography{main} 

\end{document}